\documentclass[times,11pt]{article}
\usepackage{times}

\newcommand{\nop}[1]{}

\usepackage{fancybox}
\usepackage{graphics}
\usepackage{epic}
\usepackage{eepic}
\usepackage{epsfig}
\usepackage{alltt}
\usepackage{latexsym}
\usepackage{amssymb}
\usepackage{babel}
\usepackage{theorem}

\nop{
\global\setlength\theorempreskipamount{7pt plus 2pt minus 1pt}
\global\setlength\theorempostskipamount{4pt plus 1pt minus 0.5pt}

\setlength{\textheight}{8.75in}
\setlength{\columnsep}{2.0pc}
\setlength{\textwidth}{6.8in}
\setlength{\topmargin}{0.25in}
\setlength{\headheight}{0.0in}
\setlength{\headsep}{0.0in}
\setlength{\oddsidemargin}{-.19in}
\setlength{\parindent}{1pc}

\makeatletter
\def\@normalsize{\@setsize\normalsize{12pt}\xpt\@xpt
\abovedisplayskip 10pt plus2pt minus5pt\belowdisplayskip \abovedisplayskip
\abovedisplayshortskip \z@ plus3pt\belowdisplayshortskip 6pt plus3pt
minus3pt\let\@listi\@listI} 

\def\subsize{\@setsize\subsize{12pt}\xipt\@xipt}

\def\section{\@startsection {section}{1}{\z@}{24pt plus 2pt minus 2pt}
{12pt plus 2pt minus 2pt}{\large\bf}}

\def\subsection{\@startsection {subsection}{2}{\z@}{12pt plus 2pt minus 2pt}
{12pt plus 2pt minus 2pt}{\subsize\bf}}
\makeatother
}

\newtheorem{theo}{Theorem}[section]
\newtheorem{prop}[theo]{Proposition}
\newtheorem{lem}[theo]{Lemma}
\newtheorem{coroll}[theo]{Corollary}
\newtheorem{defn}[theo]{Definition}
\newtheorem{example}[theo]{Example}

\def\endproof{\ifhmode\nobreak\qed\par\fi\medskip} 
\def\qed{\hspace*{\fill}{\rule[-0.5mm]{1.5mm}{3mm}}}

\newcommand{\Pol}{\mbox{\rm P}} 
\newcommand{\NP}{\mbox{\rm NP}}

\newcommand{\LLOG}{{{\rm L}^{\rm LOGCFL}}}

\newcommand{\LCFL}{\mbox{\rm LOGCFL}}
\newcommand{\FCFL}{\rm L^{LOGCFL}}
\newcommand{\np}{\mbox{\rm NP}} 
\newcommand{\SL}{\mbox{\rm SL}}
\newcommand{\NL}{\mbox{\rm NL}}

\newcommand{\derives}{\leftarrow} 
\renewcommand{\phi}{\varphi}
\renewcommand{\theta}{\vartheta}

\newcommand{\tuple}[1]{\langle#1\rangle}

\newif\ifpagenumbering
\pagenumberingtrue

\newcommand{\PP}{{\cal P}}

\newcommand{\BCQ}{\mbox{\rm BCQ}}

\newcommand{\BLOCKA}{{\it BLOCKA}}
\newcommand{\BLOCKB}{{\it BLOCKB}}
\newcommand{\BLOCKSA}{{\it BLOCKSA}}
\newcommand{\BLOCKSB}{{\it BLOCKSB}}
\newcommand{\BLOCKS}{{\it BLOCKS}}
\newcommand{\LINKS}{{\it LINKS}}
\newcommand{\LINK}{{\it LINK}}

\title{{\Large\bf HYPERTREE DECOMPOSITIONS AND TRACTABLE QUERIES}}

\nop{
\author{\bf Georg~Gottlob, Nicola~Leone, Francesco~Scarcello \\
Institut f\"{u}r Informationssysteme\\
        Technische Universit\"{a}t Wien \\
    Paniglgasse 16, A-1040 Wien, Austria; \\
{\tt $\{$gottlob, leone, scarcell$\}$@dbai.tuwien.ac.at}}
}

\newcommand{\DL}{{\rm L}}
\newcommand{\NC}[1]{\mbox{{\rm NC}$^#1$}}
\newcommand{\AC}[1]{\mbox{{\rm AC}$^#1$}}

\newcommand{\spath}[1]{\mbox{[$#1$]-}path}
\newcommand{\adj}[1]{\mbox{[$#1$]-}adjacent}
\newcommand{\component}[1]{\mbox{[$#1$]-}component}
\newcommand{\kdecomp}{\mbox{\tt $k$-decomp}}
\newcommand{\connected}[1]{\mbox{[$#1$]-}connected}
\newcommand{\HD}{H\!D}
\newcommand{\JT}{J\!T}

\renewcommand{\S}{{\cal S}}

\newcommand{\DB}{\mbox{\bf DB}}

\begin{document}

\setlength{\unitlength}{1mm}
\thispagestyle{empty}
 
\begin{figure}[ht]
\begin{center}
\begin{picture}(160,240)(0,0)

\put(79,165){\makebox(0,0){
\fbox{
\begin{minipage}{158mm}
\vspace*{2ex}
\begin{center}
{\LARGE
{\bf Hypertree Decompositions and Tractable Queries\\~\\}}
\vspace{0.75cm}{\bf Georg Gottlob, Nicola Leone,
Francesco Scarcello}\\
        \vspace*{0.40cm}
Institut f\"{u}r Informationssysteme \\
        Technische Universit\"{a}t Wien \\
        Paniglgasse 16, 1040 Wien \\~\\
        E-mail: $\{$gottlob,leone,scarcell$\}$@dbai.tuwien.ac.at
 
\end{center}
\end{minipage}
\vspace*{2ex}
}
}}
\put(79,110){\makebox(0,0){November 1998}}
\put(79,90){\makebox(0,0){DBAI-TR 98/21}}
\end{picture}
\end{center}
\end{figure}

\clearpage
\newpage
\addtolength{\topmargin}{1.5cm}

\author{
{\bf Georg Gottlob \quad Nicola Leone\quad
Francesco Scarcello
\thanks{Partially supported by the {\em
Istituto per la Sistemistica e l'Informatica} of the
Italian National Research Council ({\em ISI-CNR}),
under grant n.224.07.5}}
\\[1.5ex]
{
Institut f\"{u}r Informationssysteme,}\\
{
Technische Universit\"{a}t Wien} \\
{
A-1040 Wien,  Paniglgasse~16, Austria}\\[0.3ex]
{\small\tt $\{$gottlob,leone,scarcell$\}$@dbai.tuwien.ac.at}
}

\date{DBAI-TR 98/21, November 1998\\~\\}

\maketitle

\vspace{2.0cm}

\begin{abstract}
Several important decision problems on conjunctive 
queries (CQs) are $\NP$-complete in general but become 
tractable, and actually highly parallelizable,
if restricted to acyclic or nearly acyclic queries. 
Examples are the evaluation of Boolean CQs
and query containment.
These problems were shown tractable 
for conjunctive queries of 
bounded treewidth~\cite{chek-raja-97}, and of bounded
degree of cyclicity~\cite{gyss-para-84,gyss-etal-94}. 
The so far most general concept of nearly acyclic queries
was the notion of  queries of bounded query-width introduced  
by Chekuri and Rajaraman~\cite{chek-raja-97}. While CQs 
of bounded query width are tractable, it remained unclear whether 
such queries are efficiently recognizable.
Chekuri and Rajaraman~\cite{chek-raja-97} stated as an open problem whether 
for each constant $k$ it can be determined in polynomial time 
if a query has query width $\leq k$.
We give a negative answer by proving this problem
$\NP$-complete (specifically, for $k=4$).
In order to circumvent this difficulty, 
we introduce the new concept  of hypertree
decomposition of a query and the corresponding notion of 
hypertree width. We prove: (a) for each $k$, the class of 
queries with query width bounded by $k$ is properly contained 
in the class of queries whose hypertree width is bounded by $k$;
(b) unlike query width, constant hypertree-width is efficiently 
recognizable; 
(c) Boolean queries of constant hypertree width can be  efficiently
evaluated. \\
\ \\
\end{abstract}

\vspace{3.5cm}

\vfill

\thispagestyle{empty}

\newpage

\thispagestyle{empty}
\addtolength{\topmargin}{-0.75cm}

\ \\
\vspace{2.5cm}
\ \\

\tableofcontents
\newpage
\pagenumbering{arabic}

\newpage

\section{Introduction and Overview of Results}

\subsection{Conjunctive Queries}

One of the simplest but also one of the most important 
classes of database 
queries is the class of {\em conjunctive queries (CQs)}. 
In this paper we adopt the logical representation of a relational 
database~\cite{ullm-89,abit-etal-95}, 
where data tuples are identified with logical ground atoms,
and conjunctive queries are represented as datalog rules.
We will, in the first place,  deal with 
{\em Boolean} conjunctive queries (BCQs)
represented by rules whose heads are variable-free, i.e., 
propositional (see Example~\ref{ex:queries} below). 
From our results on Boolean queries, we are able to 
derive complexity results on important database problems concerning 
general (not necessarily Boolean) 
conjunctive queries.

\begin{example}
\label{ex:queries}
\em 
Consider a relational database with the following relation schemas:
{\tt
\begin{quote}
enrolled(Pers\#,Course\#,Reg\_Date)\\
teaches(Pers\#,Course\#,Assigned)\\ 
parent(Pers1, Pers2)
\end{quote}
}
The BCQ  $Q_1$ below checks whether  some
student is enrolled in a course taught by his/her parent.

\[
\!\! Q_1:\; ans \leftarrow \!\! {\tt enrolled}(S,C,R) \wedge {\tt teaches}(P,C,A)
 \wedge \; {\tt parent}(P,S).
\]

The following query $Q_2$ asks: Is there a professor who has a child
enrolled in some course? 

\[
\!\! Q_2:\; ans \leftarrow \!\! {\tt teaches}(P,C,A) \wedge {\tt enrolled}(S,C',R)
 \wedge \; {\tt parent}(P,S).
\]
\end{example}

Decision problems such as the {\em evaluation problem} of Boolean CQs,
the {\em query-of-tuple} problem (i.e., checking whether a given tuple
belongs to a CQ), 
and the {\em containment problem} for CQs have 
been studied intensively. (For recent references, see
\cite{kola-vard-98,chek-raja-97}.)
These problems -- which are all equivalent via simple logspace transformations
(see~\cite{gott-etal-98}) -- are $\NP$-complete 
in the general setting  but are polynomially solvable for a number 
of syntactically restricted subclasses. 

\subsection{Acyclic Queries and Join Trees}

Most prominent among the polynomial cases
is the class of {\em acyclic queries} 
or {\em tree queries}~\cite{yann-81,bern-good-81,good-shmu-82,yu-ozso-84,datr-mosc-86,fagi-83,fagi-etal-82,papa-yann-97}.
A query $Q$ is acyclic if its associated hypergraph
$H(Q)$ is acyclic, otherwise $Q$ is cyclic. The vertices of $H(Q)$ 
are the variables occurring in $Q$. 
Denote by $atoms(Q)$ the set of atoms in the body of $Q$, and by
$var(A)$ the variables occurring in any atom $A\in atoms(Q)$. 
The hyperedges of $H(Q)$ consist of all sets $var(A)$, 
such that $A\in atoms(Q)$. We refer to the standard notion of 
cyclicity/acyclicity in hypergraphs used in 
database theory~\cite{maie-86,ullm-89,abit-etal-95}.

A {\em join tree}
$\JT(Q)$ for a conjunctive query $Q$ is a tree
whose vertices are the atoms in the body of $Q$ such that
whenever the same variable $X$ occurs in two atoms $A_1$ and $A_2$,
then $A_1$ and $A_2$ are connected in $\JT(Q)$, and $X$ occurs in
each atom on the unique path
linking $A_1$ and $A_2$. In other words, the set of nodes in which 
$X$ occurs induces a (connected) subtree of $\JT(Q)$.
We will refer to this condition as the
{\em Connectedness Condition} of join trees.

Acyclic queries can be characterized in terms of join trees:
A query $Q$ is {\em acyclic} iff it has a join 
tree~\cite{bern-good-81,beer-etal-83}.

\begin{figure}[h]
\mbox{}\hfill\epsfig{file=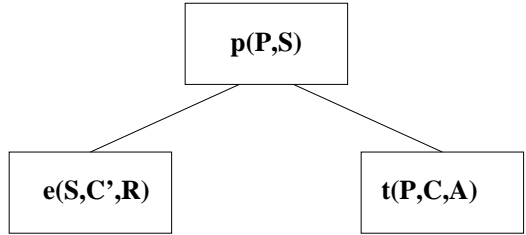,width=7cm}\hfill\mbox{}
\caption{A join tree of $Q_2$}
\label{fig:jointree-intro}
\end{figure}

\begin{example} \label{e-join-tree}
\em While query $Q_1$ of example \ref{ex:queries} is cyclic and admits
no join tree,
query $Q_2$ is acyclic. A join tree for
$Q_2$ is shown in Figure \ref{fig:jointree-intro}.
\end{example}

Acyclic conjunctive queries have highly desirable
computational properties:

(1) The problem $\BCQ$ of evaluating a Boolean conjunctive 
query can be efficiently solved if the input query is acyclic.
Yannakakis provided a (sequential) polynomial time algorithm
solving $\BCQ$ on acyclic conjunctive queries~\footnote{Note that,
since both the database $\DB$  and the query $Q$
are part of an input-instance of $\BCQ$, what we are considering
is the {\em combined complexity} of the query~\cite{vard-82}.
}~\cite{yann-81}.
The authors of the present paper have recently shown that $\BCQ$ is highly
parallelizable on acyclic queries, as it is complete
for the low complexity class $\LCFL$~\cite{gott-etal-98}.
\ (2) 
Acyclicity is efficiently recognizable, and a join tree
of an acyclic query is efficiently computable.
A linear-time algorithm for computing a
join tree is shown in \cite{tarj-yann-84};
an $\rm L^{SL}$ method has been provided in \cite{gott-etal-98}.
\ (3) The result of a (non-Boolean) acyclic conjunctive query
$Q$ can be {\em computed} in time polynomial in the combined size of
the input instance and of the output relation~\cite{yann-81}.

Intuitively, the efficient behaviour of Boolean 
acyclic queries is due to the fact
that they can be evaluated by 
processing the join tree bottom-up by performing 
upward semijoins, thus keeping
small the size of the intermediate relations (that could become
exponential if regular join were performed). This method is the 
Boolean version of Yannakakis evaluation algorithm for 
general conjunctive queries~\cite{yann-81}.

Acyclicity is a key-property responsible for the polynomial
solvability of problems that are in general $\NP$-hard 
such as $\BCQ$ \cite{chan-merl-77} and other
equivalent problems such as  Conjunctive Query 
Containment~\cite{qian-96,chek-raja-97}, Clause Subsumption, and
Constraint Satisfaction~\cite{kola-vard-98,gott-etal-98}. (For a 
survey and detailed treatment see~\cite{gott-etal-98}.)

\subsection{Queries of Bounded Width}

\begin{figure}
\mbox{}\hfill\epsfig{file=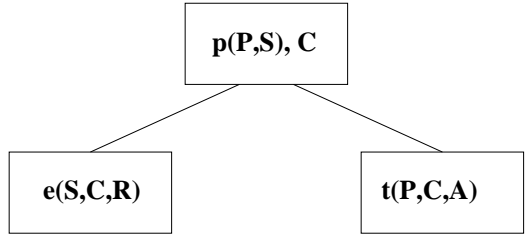,width=7cm}\hfill\mbox{}
\caption{A 2-width query decomposition of query $Q_1$}
\label{f:2width-prof-intro}
\end{figure}

The tremendous speed-up obtainable in the evaluation of acyclic queries
stimulated several research efforts towards the identification of wider classes
of queries having the same desirable properties as acyclic queries.
These studies identified a number of relevant classes
of cyclic queries which are
close to acyclic queries, because they can be decomposed
via low width decompositions to  acyclic queries.
The main classes of polynomially solvable bounded-width queries
considered in database theory and in artificial intelligence
are:
\begin{itemize}
\item {\bf The queries of bounded 
treewidth}~\cite{chek-raja-97} (see also~\cite{kola-vard-98,gott-etal-98}).
These are queries, whose variable-atom incidence graph has 
treewidth bounded by a constant.\footnote{As pointed out in 
\cite{kola-vard-98}, the notion of treewidth of a query can be 
equivalently based on the Gaifman graph of a query, i.e., the graph 
linking two variables by an edge if they occur together in a
query-atom.} The treewidth of a graph is a well-known measure of its
tree-likeness introduced by Robertson and Seymour in their work 
on graph minors~\cite{robe-seym-86}.
This notion plays a central role in algorithmic graph theory as well
as in many subdisciplines of Computer Science. We omit a formal definition.
It is well-known that checking that a graph has treewidth $\leq k$ for a
fixed constant $k$, and in the positive case, computing a $k$-width 
tree decomposition is feasible in linear time~\cite{bodl-96}.

\item {\bf  Queries of bounded degree of
cyclicity}~\cite{gyss-para-84,gyss-etal-94}. This is an interesting 
class of queries which also encompasses the class of 
acyclic queries. For space reasons, we omit a formal definition.
For each constant $k$, checking whether a query has degree of
cyclicity $\leq k$ is feasible in polynomial 
time~\cite{gyss-para-84,gyss-etal-94}.

\item  {\bf Queries of bounded query-width.}~\cite{chek-raja-97}.
This notion of bounded query-width is based on the
concept of {\em query decomposition}~\cite{chek-raja-97}.
Roughly, a query decomposition of a query $Q$ consists of a tree 
each vertex of which is labelled by a set of atoms and/or variables. 
Each variable and atom induces a connected subtree 
({\em connectedness condition}).
Each atom occurs in at least one label. The width of a query decomposition is 
the maximum of the cardinalities of its vertices. 
The {\em query width} $qw(Q)$ of $Q$ is the minimum width
over all its query decompositions.
A formal definition is given in Section~\ref{sec:query-width}; 
Figure~\ref{f:2width-prof-intro} shows a 2-width query-decomposition for
the cyclic query $Q_1$ of Example~\ref{ex:queries}.
This class is the widest of the 
three classes: Each query of bounded treewidth or of bounded 
degree of cyclicity $k$ has also bounded 
query width $k$, but for some queries the converse does not 
hold~\cite{chek-raja-97,gott-etal-98}. There are even classes of 
queries with bounded query width but unbounded treewidth.
Note, however, that no  polynomial algorithm 
for checking whether a query has width $\leq k$ was known.
\end{itemize}

All these concepts are true generalizations of the basic concept of 
acyclicity. For example, a query is acyclic iff it has query width 1.

Intuitively, a vertex of a $k$-width query decomposition
stands for the natural join of (the relations of)
its elements -- the size of this join is $O(n^k)$, where $n$ is 
the size of the input database.
Once these joins have been done,
the query decomposition can be treated exactly like a join tree
of an acyclic query, and permits to evaluate the query in time
polynomial in $n^k$~\cite{chek-raja-97}.

The problem BCQ (evaluation of Boolean conjunctive queries) 
and the bounded query-width versions of all mentioned equivalent problems, 
e.g. query-containment $Q_1\subseteq Q_2$, where the query width 
of  $Q_2$ is bounded,
can be efficiently solved 
{\em if a k-width query decomposition of the query
is given as (additional) input}. 
Chekuri and Rajamaran provided a polynomial time algorithm
for this problem~\cite{chek-raja-97}; Gottlob et
al.~\cite{gott-etal-98} later pinpointed the precise complexity of
the problem by proving it $\LCFL$-complete.

\subsection{A negative Result}

Unfortunately, unlike for acyclicity or for bounded treewidth, or 
for bounded degree of cyclicity, 
no efficient method for checking bounded query-width is known,
and a $k$-width query decomposition, which is required
for the efficient evaluation of a bounded-width query,
is not known to be polynomial time computable.

Chekuri and Rajaraman~\cite{chek-raja-97}
state this as an open problem. This problem is the first question 
we address in the present paper.

The fact that treewidth $k$ can be checked in linear time 
suggests that an analogous algorithm may work for 
query width, too. 
Chekuri and Rajaraman~\cite{chek-raja-97} express 
some optimism by writing {\em `` it would be useful to have
an  efficient algorithm that produces query decompositions of small 
width, analogous to the algorithm of Bodlaender~\cite{bodl-96} 
for decompositions of small treewidth.'' }
Kolaitis and Vardi~\cite{kola-vard-98} write somewhat more pessimistically:
{\em ``there is an important advantage of the concept 
of bounded treewidth over the concept of bounded
querywidth. Specifically, as seen 
above, the classes of structures of bounded treewidth are polynomially
recognizable, wheareas it is not known whether the same holds true for
the classes of queries of bounded querywidth''}.

Our first main result is bad news:

\vspace{0.5cm}

\centerline{\fbox{Determining whether the query width of a 
conjunctive query is at most 4 is $\NP$-complete.}}

\vspace{0.5cm}

The $\NP$-completeness proof is rather involved and will be 
given in Section~\ref{sec:np}. We give 
a rough intuition in this section. 
The proof led us to a better intuition about (i) why the problem is
$\NP$-complete, and (ii)  how this could be redressed by 
suitably modifying the notion of query width.
Very roughly, the source of $\np$-hardness can be pinpointed  as follows. 
In order to obtain a query-decomposition of width bounded by $k$, 
it can be seen 
that it is implicitly required to proceed as follows.
At any step, the decomposition is guided by a set $C$ 
of variables that still needs to be processed. Initially, i.e.,
at the root of the decomposition, $C$ consists of {\em all} variables
that occur in the query. We then choose as root of the decomposition 
tree a hypernode $R$ of $\leq k$ query 
atoms. By fixing this hypernode, we eliminate a set of variables,
namely those which occur in the atoms of $R$. The remaining set 
of variables disintegrates into connected components.
We now expand the decomposition tree by attaching 
children, and thus, in the long run, subtrees to $R$.
It can be seen that each subtree rooted in $R$ must correspond to 
one or more of the connected components and each component occurs 
in exactly one subtree (otherwise the connectedness condition would be
violated). In particular, since each atom should be eventually 
covered, each remaining atom must be covered by some
subtree, i.e., must occur in some subtree. 
This process goes on until all variables are eliminated and until 
all query atoms are eventually covered.
The definition of query width requires that this  covering be {\em exact},
i.e., that each atom containing a variable of a certain component $C$
occurs only in the subtree corresponding to $C$. 
(Again, the requirement of exact covering is due to the connectedness 
condition.) The $\NP$ hardness is due to this requirement
of {\em exact} covering. In our $\NP$-completeness proof, we 
thus tried to reduce the problem of {\tt EXACT COVERING BY 3SETS} to the 
query width problem. And this attempt was successful.

\subsection{Hypertree Decompositions: Positive Results}

To circumvent the high complexity of query-decompositions, 
we introduce a new concept of
decomposition, which we call {\em hypertree decomposition}. 
The definition of  hypertree decomposition (see Section~\ref{sec:hyper})
uses a more liberal notion of covering.
When choosing a set $SC$ of components to be processed, 
      the variables of $SC$  are no longer required 
      to exactly {\em coincide} with the variables occurring in the labels 
      of the decomposition-subtree corresponding to $SC$,
      but it is sufficient that the former be a subset of the latter.
Based on this more liberal notion of decomposition,
we define the corresponding notion of {\em hypertree width} 
in analogy to the concept of query width. 

We denote the query width of a query by $qw(Q)$ and its 
hypertree width by $hw(Q)$.
We are able to prove the following results:

\begin{center}
\fbox{\begin{minipage}{15.8cm}
{\em 
\begin{enumerate}
\item For each conjunctive query $Q$ it holds that $hw(Q)\leq qw(Q)$.
\item There exist queries $Q$ such that $hw(Q)<qw(Q)$.
\item For each fixed constant $k$, the problems
of determining whether $hw(Q)\leq k$ and of computing (in the
positive case) a hypertree decomposition of width $\leq k$ 
are feasible in polynomial time.
\item For fixed $k$, evaluating a Boolean conjunctive query $Q$ with
$hw(Q)\leq k$ is feasible in polynomial time. 
\item The result of a (non-Boolean) conjunctive query
$Q$ of bounded hypertree-width
can be {\em computed} in time polynomial in the combined size of
the input instance and of the output relation.
\item Tasks 3 and 4 are not only polynomial, but are highly 
parallelizable. In particular, for fixed $k$, 
checking whether $hw(Q)\leq k$ is in the parallel
complexity class $\LCFL$; computing a 
hypertree decomposition of width $k$ (if any) is in functional $\LCFL$, i.e., 
is feasible by a logspace transducer that uses an oracle in $\LCFL$;
evaluating $Q$ where $hw(Q)\leq k$
on a database is complete for $\LCFL$ under logspace 
reductions.
\end{enumerate}
}
\end{minipage}}
\end{center}

Similar results hold for 
the equivalent problem of conjunctive query containment $Q_1\subseteq
Q_2$, where $hw(Q_2)\leq k$, and for all other of the aforementioned 
equivalent problems.

Let us comment on these results. By statements 1 and 2, the concept 
of hypertree width is a proper generalization of the notion of query
width. By statement 3, constant hypertree-width is efficiently checkable,
and by statement 4, queries of constant hypertree width can be  efficiently
evaluated. In summary, this is truly good news. It means that the
notion of constant hypertree width not only shares the desirable properties of 
constant query-width, it also does {\em not} share the bad properties of
the latter, and, in addition, is  a more general concept. 

It thus turns out that the high complexity of determining constant query width 
is not, as one would usually expect,  the price for the generality 
of the concept. Rather, it is due to some peculiarity in its definition
related to the exact covering paradigm. In the definition of hypertree
width we succeeded to eliminate these problems without paying any 
additional charge, i.e., hypertree width comes as a freebie!

Statement 6 asserts  that the main algorithmic tasks related to 
constant hypertree-width are in the very low complexity class
$\LCFL$. 
This class consists of all decision problems that are logspace reducible
to a context-free language.
An obvious example of a problem complete for \LCFL\ is Greibach's
hardest context-free language~\cite{grei-73}. There is a number of
very interesting natural problems known to be \LCFL-complete (see,
e.g.~\cite{sudb-77,skyu-vali-85,gott-etal-98}). 
The relationship between \LCFL\ and other well-known 
complexity classes is summarized in the following
chain of inclusions:

$$\AC{0}\subseteq \NC{1}\subseteq \DL \subseteq \SL\subseteq \NL \subseteq
\LCFL\subseteq \AC{1}\subseteq  \NC{2}\subseteq \Pol \subseteq \NP$$

Here $\DL$ denotes logspace, 
$\AC{i}$ and $\NC{i}$ are {\em logspace-uniform} classes
based on the corresponding types of Boolean circuits, 
$\SL$ denotes symmetric logspace, $\NL$ denotes nondeterministic 
logspace, \Pol\ is polynomial time, and $\NP$ is nondeterministic 
polynomial time. For the definitions of all these classes,
and for references concerning their mutual relationships,
see~\cite{john-90}. 

Since 
$\LCFL\subseteq \AC{1}\subseteq \NC{2}$, the problems in \LCFL\ are
all highly parallelizable. In fact, they are solvable in logarithmic
time by a CRCW PRAM with a polynomial number of processors, or in
$\log^2$-time by an EREW PRAM with a polynomial number of processors.

\subsection{Structure of the Paper}

Basic notions of database and complexity theory are 
given in Section~2.
Section~3  
deals with query decompositions and includes
the $\NP$-completeness proof for the problem of
deciding bounded query-width.
The new  notions of hypertree decomposition and hypertree width are
formally defined in Section~4, where also some examples are given,
and it is shown that bounded hypertree-width queries are efficiently
evaluable. 
In Section 5, 
the alternating algorithm 
{\tt k-decomp} that checks  whether a query has hypertree width $\leq k$
is presented. 
This algorithm is shown to run on a logspace ATM having 
polynomially sized accepting computation trees, thus the problem is 
actually in $\LCFL$. Finally, a short sketch of 
a deterministic polynomial algorithm (in form of a datalog program)
for checking  whether a query has hypertree width $\leq k$
is given.
In Section~6 the notion 
of hypertree decomposition is shown to be the most general 
among the most important
related notions, e.g., the notion of query decomposition.

\section{Preliminaries}

\subsection{Databases and Queries}

For a background on databases, conjunctive queries, etc., 
see~\cite{ullm-89,abit-etal-95,maie-86}. 
We define only the most relevant concepts here.

A relation schema $R$ consists of a name (name of the relation) $r$
and a finite ordered list of attributes.
To each attribute $A$ of the schema, 
a countable domain $Dom(A)$ of atomic values
is associated. 
A {\em relation instance} (or simply, a {\em relation}) 
over schema $R=(A_1,\ldots,A_k)$ is a finite subset 
of the cartesian product $Dom(A_1)\times\cdots\times Dom(A_k)$. 
The elements of relations are called {\em tuples}.
A database schema $DS$ consists of a finite set of 
relation schemas. A {\em database instance}, or simply {\em database},
$\DB$ over database schema $DS=\{R_1,\ldots,R_m\}$ consists of relation
instances $r_1,\ldots,r_m$ for the schemas $R_1,\ldots, R_m$,
respectively, and a finite universe 
$U\subseteq 
\bigcup_{R_i(A^i_1,\ldots,A^i_{k_i})\in DS} 
       (Dom(A^i_1)\cup\cdots\cup Dom(A^i_{k_i}))$ 
such that all data values occurring in $\DB$ are from $U$.

In this paper we will adopt 
the standard convention~\cite{abit-etal-95,ullm-89} of  identifying a relational 
database instance with a logical theory consisting of 
ground facts. Thus, a tuple $\langle a_1,\ldots
a_k\rangle$, belonging to relation $r$, will be identified with 
the ground atom $r(a_1,\ldots,a_k)$. 
The fact that a tuple 
$\langle a_1,\ldots,a_k\rangle$ belongs to relation $r$ of 
a database instance $\DB$ is thus simply denoted by 
$r(a_1,\ldots,a_k)\in {\DB}$.

A (rule based) {\em conjunctive query} $Q$ on a database schema 
$DS=\{R_1,\ldots,R_m\}$ consists 
of a rule of the form 
$$Q:\ \  ans({\bf u})\leftarrow r_1({\bf u_1})\wedge\cdots\wedge r_n({\bf u_n}), $$
where $n \geq 0$, $r_1,\dots, r_n$ are relation names (not necessarily
distinct) of $DS$; $ans$ is a relation name not in $DS$; and 
${\bf u, u_1,\ldots, u_n}$ are lists of terms (i.e., variables or
constants) of appropriate length. The set of variables occurring 
in $Q$ is denoted by $var(Q)$. The set of atoms contained in the 
body of $Q$ is referred to as $atoms(Q)$. 
Similarly, for any atom $A\in atoms(Q)$, $var(A)$ denotes the set of
variables occurring in $A$; and for a set of atoms $R\subseteq
atoms(Q)$, define $var(R)= \bigcup_{A\in R} var(A)$.

The {\em answer} of $Q$ on a database instance $\DB$ 
with associated universe $U$, consists of 
a relation $ans$ whose arity is equal to the length of ${\bf u}$,
defined as follows. $ans$ contains all tuples $ans({\bf u})\theta$ such 
that $\theta:var(Q)\longrightarrow U$ is a substitution 
replacing each variable in $var(Q)$ by a value of $U$ and such that 
for $1\leq i\leq n$, $r_i({\bf u_i})\theta\in {\DB}$. (For an atom
$A$, $A\theta$ denotes the atom obtained from $A$ by uniformly 
substituting $\theta(X)$ for each variable $X$ occurring in $A$.)

The conjunctive query $Q$ is a  {\em Boolean conjunctive query (BCQ)} 
if its head atom $ans({\bf u})$ 
does not contain  variables
and is thus a purely
propositional atom. $Q$ evaluates to {\em true} if there exists 
a substitution $\theta$ such that 
for $1\leq i\leq n$, $r_i({\bf u_i})\theta\in {\DB}$; otherwise 
the query evaluates to {\em false}. 

The head literal in Boolean conjunctive queries is actually 
inessential, therefore we may omit it when specifying a 
Boolean conjunctive query.

Note that conjunctive queries as defined here correspond to 
conjunctive queries in the more classical setting of relational 
calculus, as well as to 
SELECT-PROJECT-JOIN queries in the classical setting of relational 
algebra, or to simple SQL queries of the type 
$${\tt SELECT\ } R_{i_1}.A_{j_1},\ldots R_{i_k}.A_{j_k} {\tt \ FROM\ }
R_1,\ldots R_n {\tt \ WHERE\ }
cond,$$ such that $cond$ is a conjunction of conditions of the form 
$R_i.A=R_j.B$ or $R_i.A=c$, where $c$ is a constant.

A query $Q$ is {\em acyclic}~\cite{beer-etal-83,bern-good-81} 
if its associated hypergraph
$H(Q)$ is acyclic, otherwise $Q$ is cyclic. The vertices of $H(Q)$ 
are the variables occurring in $Q$. 
Denote by $atoms(Q)$ the set of atoms in the body of $Q$, and by
$var(A)$ the variables occurring in any atom $A\in atoms(Q)$. 
The hyperedges of $H(Q)$ consist of all sets $var(A)$, 
such that $A\in atoms(Q)$. We refer to the standard notion of 
cyclicity/acyclicity in hypergraphs used in 
database theory~\cite{maie-86,ullm-89,abit-etal-95}.

A {\em join tree}
$\JT(Q)$ for a conjunctive query $Q$ is a tree
whose vertices are the atoms in the body of $Q$ such that
whenever the same variable $X$ occurs in two atoms $A_1$ and $A_2$,
then $A_1$ and $A_2$ are connected in $\JT(Q)$, and $X$ occurs in
each atom on the unique path
linking $A_1$ and $A_2$. In other words, the set of nodes in which 
$X$ occurs induces a (connected) subtree of $\JT(Q)$
({\em connectedness condition}).

Acyclic queries can be characterized in terms of join trees:
A query $Q$ is {\em acyclic} iff it has a join 
tree~\cite{bern-good-81,beer-etal-83}.

\begin{figure}[h]
\mbox{}\hfill
\epsfig{file=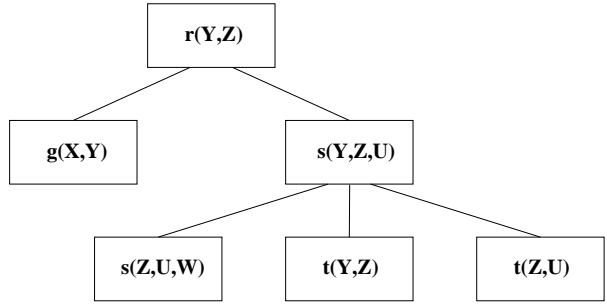,width=8cm}\hfill\mbox{}
\caption{A join tree of $Q_3$}
\label{fig:jointree}
\end{figure}

\begin{example} \label{e-join-tree-b}
\em While query $Q_1$ of example \ref{ex:queries} is cyclic and admits
no join tree, query $Q_2$ is acyclic. A join tree for
$Q_2$ is shown in Figure \ref{fig:jointree-intro}.

Consider the following query $Q_3$:
$$ ans \leftarrow
   r(Y,Z) \wedge g(X,Y) \wedge s(Y,Z,U)\wedge s(Z,U,W)
   \wedge t(Y,Z)\wedge t(Z,U)$$
A join tree for $Q_3$ is shown in Figure~\ref{fig:jointree}.
\end{example}

Acyclic conjunctive queries have highly desirable
computational properties:
\begin{enumerate}
\item The problem $\BCQ$ of evaluating a Boolean conjunctive 
query can be efficiently solved if the input query is acyclic.
Yannakakis provided a (sequential) polynomial time algorithm
solving $\BCQ$ on acyclic conjunctive queries~\footnote{Note that,
since both the database $\DB$  and the query $Q$
are part of an input-instance of $\BCQ$, what we are considering
is the {\em combined complexity} of the query~\cite{vard-82}.
}~\cite{yann-81}.
The authors of the present paper have recently shown that $\BCQ$ is highly
parallelizable on acyclic queries, as it is complete
for the low complexity class $\LCFL$~\cite{gott-etal-98}.
\item
Acyclicity is efficiently recognizable, and a join tree
of an acyclic query is efficiently computable.
A linear-time algorithm for computing a
join tree is shown in \cite{tarj-yann-84};
an $\rm L^{SL}$ method has been provided in \cite{gott-etal-98}.
\item The result of a (non-Boolean) acyclic conjunctive query
$Q$ can be {\em computed} in time polynomial in the combined size of
the input instance and of the output relation~\cite{yann-81}.
\end{enumerate}

Intuitively, the efficient behaviour of Boolean 
acyclic queries is due to the fact
that they can be evaluated by 
processing the join tree bottom-up by performing 
upward semijoins, thus keeping
small the size of the intermediate relations (that could become
exponential if regular join were performed). This method is the 
Boolean version of Yannakakis evaluation algorithm for 
general conjunctive queries~\cite{yann-81}.

Acyclicity is a key-property responsible for the polynomial
solvability of problems that are in general $\NP$-hard 
such as $\BCQ$ \cite{chan-merl-77} and other
equivalent problems such as  Conjunctive Query 
Containment~\cite{qian-96,chek-raja-97}, Clause Subsumption, and
Constraint Satisfaction~\cite{kola-vard-98,gott-etal-98}. (For a 
survey and detailed treatment see~\cite{gott-etal-98}.)

\subsection{The class LOGCFL}\label{sec:logcfl}

\LCFL\ consists of all decision problems that are logspace reducible
to a context-free language.
An obvious example of a problem complete for \LCFL\ is Greibach's
hardest context-free language~\cite{grei-73}. There are a number of
very interesting natural problems known to be \LCFL-complete (see,
e.g.~\cite{gott-etal-98,sudb-77,skyu-vali-85}). 
The relationship between \LCFL\ and other well-known 
complexity classes is summarized in the following
chain of inclusions:

$$\AC{0}\subseteq \NC{1}\subseteq \DL \subseteq \SL\subseteq \NL \subseteq
\LCFL\subseteq \AC{1}\subseteq  \NC{2}\subseteq \Pol \subseteq \NP$$

Here $\DL$ denotes logspace, 
$\AC{i}$ and $\NC{i}$ are {\em logspace-uniform} classes
based on the corresponding types of Boolean circuits, 
$\SL$ denotes symmetric logspace, $\NL$ denotes nondeterministic 
logspace, \Pol\ is polynomial time, and $\NP$ is nondeterministic 
polynomial time. For the definitions of all these classes,
and for references concerning their mutual relationships,
see~\cite{john-90}.

Since -- as mentioned in the introduction -- 
$\LCFL\subseteq \AC{1}\subseteq \NC{2}$, the problems in \LCFL\ are
all highly parallelizable. In fact, they are solvable in logarithmic
time by a CRCW PRAM with a polynomial number of processors, or in
$\log^2$-time by an EREW PRAM with a polynomial number of processors.

In this paper, we will use an important characterization of \LCFL\  
by Alternating Turing Machines. We assume that the reader is 
familiar with 
the {\em alternating Turing machine (ATM)} computational model
introduced by Chandra, Kozen, and Stockmeyer \cite{chan-etal-81}.
Here we assume w.l.o.g. that the states of an ATM are partitioned 
into existential and universal states.

As in~\cite{ruzz-80}, we define a 
{\em computation tree} of an ATM $M$ on a input string $w$ as a tree
whose nodes are labeled with configurations of $M$ on $w$, 
such that the descendants of any non-leaf labeled by a universal 
(existential) configuration include all (resp. one) of the successors
of that configuration. A computation tree is {\em accepting} if the
root is labeled with 
the initial configuration, and all the leaves are accepting configurations.

\nop{
Note that, since existential nodes have exactly one successor
in a computation tree, their quality is determined by Rule 4 above;
in particular, the quality of an internal node representing an
existential configuration is Accept (Reject) if its successor
configuration has quality Accept (resp. Reject). 

A computation tree for an ATM $M$ is an {\em accepting (computation)
tree} if its root has the quality Accept.}

Thus, an accepting tree yields a certificate that the input is accepted.
A complexity measure considered by Ruzzo \cite{ruzz-80} for the
alternating Turing machine is the tree-size, i.e. the
minimal size of an accepting computation tree.

\begin{defn}[\cite{ruzz-80}]
A decision problem $\PP$ is solved by an alternating Turing machine $M$
within {\em simultaneous} tree-size and space bounds $Z(n)$ and $S(n)$
if, for every ``yes'' instance $w$ of $\PP$, there is at least one accepting
computation tree for $M$ on $w$ of size (number of nodes) $\leq Z(n)$,
each node of which represents a configuration using space $\leq S(n)$,
where $n$ is the size of $w$.
(Further, for any ``no'' instance $w$ of $\PP$ there is no accepting 
computation tree for $M$.)
\end{defn}

Ruzzo \cite{ruzz-80} proved the following important characterization 
of \LCFL\ :

\begin{prop}[Ruzzo~\cite{ruzz-80}]
\label{prop:atm}
$\LCFL$  coincides with the class of all decision problems
recognized by
ATMs operating simultaneously in tree-size
$O(n^{O(1)})$ and space $O(\log n)$.
\end{prop}

\section{Query Decompositions}

\subsection{Bounded Query Width and Bounded Query Decompositions}
\label{sec:query-width}

The following definition of query decomposition 
is a slight modification of the original
definition given by Chekuri and Rajaraman~\cite{chek-raja-97}.
Our definition is a bit more liberal because, for any conjunctive query $Q$, 
we do not take care of the atom $head(Q)$, as well as of the
constants possibly occurring in $Q$. 
However, in this paper, 
we will only deal with Boolean conjunctive queries without
constants, for which the two notions coincide.

\begin{defn}\label{def:query-width}
\em
A {\em query decomposition} of a conjunctive query $Q$ is 
a pair $\tuple{T,\lambda}$, where 
$T=(N,E)$ is a tree,  and $\lambda$ is a labeling function which associates
to each vertex $p\in N$ a set $\lambda(p)\subseteq (atoms(Q)\cup var(Q))$,
such that the following conditions are satisfied:
\begin{enumerate}
\item for each atom $A$ of $Q$, there exists $p\in N$ 
      such that $A\in \lambda(p)$;
\item for each atom $A$ of $Q$, the set $\{p\in N\; |\; A\in \lambda(p)\}$
      induces a (connected) subtree of $T$;
\item for each variable $Y\in var(Q)$, the set
   $$\{ p\in N\; |\; Y\in \lambda(p)\} \cup 
     \{ p\in N\; |\; Y \mbox{ occurs in some atom } A\in \lambda(p)\}$$
    induces a (connected) subtree of $T$.
\end{enumerate}

The {\em width} of the query decomposition $\tuple{T,\lambda}$
is $max_{p\in N} |\lambda(p)|$.
The {\em query width} $qw(Q)$ of $Q$ is the minimum width 
over all its query decompositions.
A query decomposition for $Q$ is {\em pure}
if, for each vertex $p\in N$, $\lambda(p)\subseteq atoms(Q)$.
\end{defn}
Note that Condition~3 above is the analogue of the
connectedness condition of join trees and thus 
we will refer to it as the {\em Connectedness Condition}, as well.

\begin{figure}
\mbox{}\hfill\epsfig{file=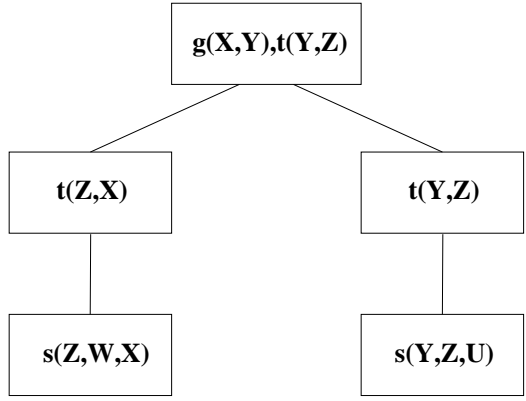,width=7cm}\hfill\mbox{}
\caption{A 2-width query decomposition of query $Q_4$}
\label{f:2width-prof}
\end{figure}

\begin{example}\label{ex:query-width}
\em
Figure~\ref{f:2width-prof-intro} shows a 2-width query decomposition
for the cyclic query of Example~\ref{ex:queries}.

Consider the following query $Q_4$:
$$ ans \leftarrow
   s(Y,Z,U) \wedge g(X,Y) \wedge t(Z,X)\wedge s(Z,W,X)\wedge t(Y,Z)$$
$Q_4$ is a cyclic query, and its query width equals 2.
A 2-width decomposition of $Q_4$ is shown in
Figure~\ref{f:2width-prof}. Note that this query decomposition is pure.
\end{example}

The next proposition, which is proved elsewhere~\cite{gott-etal-98},
shows that we can focus our attention
on pure query decompositions.
\begin{prop}[\cite{gott-etal-98}]\label{prop:pure-dec}
Let $Q$ be a conjunctive query and $\tuple{T,\lambda}$
a $c$-width query decomposition of $Q$.
Then,
\begin{enumerate}
\item
there exists 
a pure $c$-width query decomposition $\tuple{T,\lambda'}$ of $Q$;
\item
$\tuple{T,\lambda'}$ is logspace computable from $\tuple{T,\lambda}$.
\end{enumerate}
\end{prop} 

Thus, by Proposition~\ref{prop:pure-dec}, for any conjunctive query
$Q$, $qw(Q)\leq k$ if and only if $Q$ has a pure $c$-width
decomposition, for some $c\leq k$.

{\em k-bounded-width queries} 
are queries whose query width is bounded by a fixed
constant $k>0$.
The notion of bounded query-width generalizes the notion of
acyclicity~\cite{chek-raja-97}. Indeed, acyclic queries are
exactly the conjunctive queries of query width 1, because any join
tree is a query decomposition of width 1.

Bounded-width queries share an important computational property
with acyclic queries:
$\BCQ$ can be efficiently solved on queries of k-bounded
query-width, if a k-width query decomposition of the query
is given as (additional) input.
Chekuri and Rajamaran provided a polynomial time algorithm
for this problem~\cite{chek-raja-97}; while Gottlob et al. pinpointed
that the precise complexity of the problem is $\LCFL$.

Unfortunately, different from acyclicity,
no efficient method for checking bounded query-width is known.
In fact, we next prove that deciding whether a conjunctive query
has a bounded-width query decomposition is $\NP$-complete.

\subsection{Recognizing bounded query-width is  $\NP$-complete}
\label{sec:np}

A {\em $k$-element-vertex} of a query decomposition $(T,\lambda)$ 
is a vertex $v$ of $T$ 
such that $|\lambda(v)|=k$. 

\begin{lem}
\label{lem:blocks}
Let $Q$ be a query having variable  set $var(Q)= \Gamma\cup Rest$, where
$$\Gamma=\{V_{ij}\,|\, 1\leq i< j\leq 8\},$$
and $Rest$ is an arbitrary set of 
further variables. Assume the set $atoms(Q)$ contains as subset a set $\Pi=\{P_1,\ldots,P_8\}$  
of 8 atoms, 
where, for $1\leq i\leq 8$, $var(P_i)\cap \Gamma=\{V_{1i},V_{2i},\ldots,V_{i-1\,i},V_{i\,i+1},
\ldots,V_{i8}\}$, i.e., 
$$var(P_i)\cap \Gamma=\bigcup_{k<i}\{V_{ki}\}\  \cup\  \bigcup_{i<k}\{V_{ik}\}.$$
and 
$\forall A\in atoms(Q)-\Pi:\, var(A)\cap\Gamma=\emptyset$.

If $Q$ admits a pure query decomposition $(T,\lambda)$ of width 4, then
there exist two adjacent $4$-element-vertices $p_1$ and $p_2$ of $T$ such that 
$\lambda(p_1)\cup \lambda(p_2)=\Pi$.
\end{lem}

\begin{proof} 
Assume this doesn't hold. Let $v$ be any  vertex from $T$ such that 
$\Gamma\cap var(v)\neq\emptyset$. Let $R=\Pi-\lambda(v)$.
The atoms of $R$ must occur somewhere in the tree $T$. By the {\em
connectedness condition}, all atoms of $R$ must occur in the labels of 
some neighbours of $v$. Moreover, by our assumption, the atoms of 
$R$ are not contained in the label of a single neighbour of $v$.
Thus, there exist two neighbours $v_1,v_2$ of $v$ and two different
atoms 
$P_i\,P_j\in R$ such that $P_i\in\lambda(v_1)-\lambda(v_2)$ 
and $P_j\in\lambda(v_2)-\lambda(v_1)$. Assume, w.l.o.g., that $i<j$.
Then $V_{ij}\in var(v_1)$, $V_{ij}\in var(v_2)$, but $V_{ij}\not\in
var(v)$. This, however, violates the {\em connectedness condition}.
Contradiction.
\end{proof}

\begin{defn} \em
Let $S$ be a set of $n$ elements. A {\em 3-partition}  $\{S_a,S_b,S_c\}$ of $S$ consists of  
three nonempty subsets $S_a,S_b,S_c\subset S$ such that $S_a\cup S_b\cup S_c=S$,
and $S_x\cap S_y=\emptyset$ for $x\neq y$ from $\{a,b,c\}$.
The sets $S_a,S_b,S_c$ are referred to as {\em classes}.

A {3-Partitioning-System} (short {\em 3PS}) $\Sigma$ on a base set $S$ 
is a set of 3-partitions 
of $S$:
$$\Sigma\, =\,  \{\ \,\{S_a^1,S_b^1,S_c^1\},   \{S_a^2,S_b^2,S_c^2\}, \ldots,  
\{S_a^m,S_b^m,S_c^m\}\ \},$$
where $\forall \sigma,\sigma'\in \Sigma$: 
$\sigma\neq \sigma' \Rightarrow \sigma\cap\sigma'=\emptyset$ 
(i.e., no class occurs in two or more elements of $\Sigma$). 

We define $classes(\Sigma):=\bigcup_{\sigma\in \Sigma} \sigma$.
The base set $S$ of $\Sigma$ is referred-to as $base(\Sigma)$: 
$base(\Sigma) =\bigcup_{C\in classes(\Sigma)}C$.

A 3PS is  {\em strict} if 
for all $S',S'', S'''\in classes(\Sigma)$ 
either $\{S',S'',S'''\}=\sigma$ for some 
$\sigma\in\Sigma$ or $S'\cup\S''\cup S'''\subset S$. 
In other words: the only way to 
obtain $S$ as a union of three classes is via the specified 3-partitions of $\Sigma$;
any other union of three classes results in a proper subset of $S$.

A 3PS $\Sigma$ is referred to as an {\em $(m,k)$-3PS} if 
$|\Sigma|\geq m$ and $\forall C\in classes(\Sigma): |C|\geq k$.
\end{defn}

\begin{lem}
\label{lem:dirty}
For each $m>0$ and $k>0$ a strict $(m,k)$-3PS 
can be computed in polynomial time.
\end{lem}
\begin{proof}  
Fix $m$ and $k$. We will construct a set $S$ of the desired cardinality $n\leq 
27m^3 + 2m+ 3k$
such that there exists an $(m,k)$-3PS for $S$. 

In order to construct $S$, first start with a set $S_0$ of $2m$ elements.
It is a trivial combinatorial fact that we can choose at least $m$ different
3-partitions of $S_0$ (recall that $m>3$, thus the actual number 
of 3-partitions we can build is much higher). Thus, let $\Sigma_0$ be a 
(not necessarily strict) 3PS on base set $S_0$, such that 
$|\Sigma_0|=m$. We now basically
have to achieve two goals: (1) we have to transform $\Sigma_0$ 
to a {\em strict} 3PS, and,  (2) we have to make sure that all classes have at least 
$k$ elements.

In order to achieve goal (1), let $New$ 
be a set of $27m^3$ fresh elements. Enumerate (e.g., in a FOR loop) all 
combinations of three sets $S_x^i,S_y^j,S_z^\ell\in classes(\Sigma_0)$ and 
check for each such triplet, whether it violates the strictness-condition.
If so, choose a fresh (i.e., so far unused) element $a$ from $New$, insert it into
$S_0$ and use it as follows to redress the situation: For any set $\sigma\in \Sigma_0$ 
in which neither of $S_x^i,S_y^j,S_z^\ell$ occurs, insert $a$ into exactly one of the three 
classes (which one is irrelevant). 
This means that the so augmented $\sigma$ is a partition of the new $S_0$.
On the other hand, for each $\sigma\in\Sigma_0$ 
such that $\sigma \cap \{S_x^i,S_y^j,S_z^\ell\} \neq \emptyset$,
choose a class $C\in\sigma$ such that 
$C\not\in \{S_x^i,S_y^j,S_z^\ell\}$ and insert 
$a$ into $C$. 
It follows that the triplet 
$S_x^i,S_y^j,S_z^{\ell}$ 
no longer violates the strictness condition w.r.t. the new 
set $S_0$ (containing $a$), because $a$ is not in the union 
$S_x^i\cup S_y^j\cup S_z^\ell$, and thus this union is a strict
subset of $S_0$. Note that this operation does never invalidate the 
strictness condition of any other triplet. After we have repeated this 
procedure for each triplet, we end-up with a strict 3PS for the resulting
set $S_0$. Call this resulting set $S^+$ and denote the resulting 
strict 3PS by $\Sigma^+$. 
Retain that the  set $S^+$ has 
less than $2m+(3m)^3=2m+27m^3$ elements. 

In order to achieve goal (2), we simply add  a set $New'$ 
of $3k$ further fresh elements to 
$S^+$, obtaining $S^*= S^+\cup New'$. We furthermore 
partition $New'$ into three sets $O_1$, $O_2$, and $O_3$ 
of equal cardinality $k$ 
and do the following for each 3-partition 
$\sigma=\{C_a,C_b,C_c\}$ of $\Sigma^+$,
where $\langle C_a,C_b,C_c\rangle$ is any arbitrarily chosen order of the 
elements of $\sigma$: 
perform $C_a:= C_a\cup O_1$ and $C_b:= C_b\cup O_2$ and 
$C_c:= C_c\cup O_3$. 

The resulting 3PS $\Sigma$ is strict, has 
$|\Sigma|=m$, $|base(\Sigma)|\leq 27m^3 + 2m+ 3k$,
and each class $C$ of $classes(\Sigma)$ 
has $|C|\geq k$. We are done. Note that our method for generating a strict
3PS works in polynomial time. 
\end{proof}

\begin{theo}
Deciding whether the query width of 
a conjunctive query is at most $4$ is \NP-complete.
\end{theo}

\begin{proof} 
{\em 1. Membership.} 
It is easy to see that if there exists a query decomposition
of width bounded by $4$, then there also exists one of polynomial size
(in fact, by a simple restructuring technique we can always 
remove identically labeled vertices from a decomposition tree, and thus 
for any conjunctive query $Q$
only  $O(|atoms(Q)\cup var(Q)|^4)$  need to be considered). 
Therefore, a query decomposition
of width $\leq k$ can be found by 
a nondeterministic guess followed by a polynomial 
correctness check. The problem is thus in \NP. 

{\em 2. Hardness.} We transform the well-known \NP-complete problem 
{\tt EXACT COVER BY 3-SETS } ({\tt XC3C})~\cite{gare-john-79} 
to the problem of deciding whether, for  a conjunctive query $Q$,
$qw(Q)\leq 4$ holds.
An instance of {\tt EXACT COVER BY 3-SETS} consists of a pair  $I=(R,\Delta)$ 
where $R$ is a set of $r=3s$ elements, and $\Delta$ is a collection of 
$m$ 3-element subsets of $R$. The question is whether we can select 
$s$ subsets out of $\Delta$ such that they form a partition of $R$.

Consider an instance $I=(R,\Delta)$ of {\tt XC3S}. 
Let $\Delta=\{D_i|1\leq i\leq m\}$ 
and let $D_i=\{X_a^i,X_b^i,X_c^i \}$ for $1\leq i\leq m$ (note that for $i\neq j$,
some $X_\alpha^i$ and $Y_\beta^j$ may coincide). 

Generate a strict $(m+1,2)$ 3PS 
$\Sigma=\{\sigma_0,\sigma_1,\ldots,\sigma_m\}$ on some 
base set $S=base(\Sigma)$.  By 
Lemma~\ref{lem:dirty}, this can be done in polynomial time. 
Let $\sigma_i=\{S_a^i,S_b^i,S_c^i\}$ for $0\leq i\leq m$.

Identify each element of $S$ with a separate variable and establish 
a fixed precedence order $\prec$ among the elements (variables) of $S$.
If $S'$ is a subset of $S$, and $S'=\{Z_1,\ldots,Z_l\}$, where 
$Z_1\prec Z_2\cdots\prec Z_l$, then we will abbreviate the list of 
variables $Z_1,\ldots,Z_l$ by $S'$ in query atoms. For example, instead
of writing $p(a,Z_1,\ldots,Z_l,b)$, we write $p(a,S',b)$.

In order to transform the given instance $I=(R,\Delta)$ of 
{\tt XC3S} to a conjunctive query $Q$, let us first 
define the following sets of variables $\Gamma^\ell$ and 
$\Pi^\ell_i$ , which are all taken to be disjoint  
from the variables in $S$.

For $0\leq \ell\leq s$, let $$\Gamma^\ell=\{ V_{ij}^\ell\,|\, 1\leq i<j\leq 8  \},$$
and for ($1\leq i\leq 8$), let 

$$\Pi_i^\ell=\{V_{1i}^\ell,V_{2i}^\ell,\ldots,V_{i-1\,i}^\ell,V_{i\,i+1}^\ell ,\ldots,V_{i8}^\ell\}.$$

Let $S'_a$ and $S''_a$ be two nonempty sets which partition $\S^0_a$.
(Such a partition exists because $S^0_a$ contains
at least two elements.) 

Define, for $0\leq \ell\leq s$  the following sets of query atoms:

$$ \BLOCKA^\ell=\{q(\Pi_1^\ell,S'_a,Z_\ell),\, p_a(\Pi_2^\ell,S''_a),\, 
p_b(\Pi_3^\ell,S_b^0),\, p_c(\Pi_4^\ell,S_c^0) \}$$

$$ \BLOCKB^\ell=\{q(\Pi_5^\ell,S'_a,Y_\ell),\, p_a(\Pi_6^\ell,S''_a),\, p_b(\Pi_7^\ell,S_b^0),\, p_c(\Pi_8^\ell,S_c^0) \},$$ 
where the $Y_\ell$ and $Z_\ell$ variables are distinct fresh variables 
not occurring in any previously defined set.
We further define 
$$\BLOCKSA =\bigcup_{0\leq \ell\leq s} \BLOCKA^\ell,\ \;\;\;
\BLOCKSB =\bigcup_{0\leq \ell\leq s} \BLOCKB^\ell,$$
$$\ \  {\rm and}\ \ \BLOCKS = \; \BLOCKSA \; \cup\; \BLOCKSB.$$

Define, for $1\leq \ell \leq s$:

$$\LINK_\ell = \{ link(Y_{\ell-1},Z_\ell) \} \ \ \  {\rm and} \ \ \ \LINKS=\bigcup_{1\leq\ell\leq s}\LINK_\ell.$$

Finally, define for each set $D_i=\{X_a^i,X_b^i,X_c^i\}$ of $\Delta$, $1\leq i\leq m$, 
the set of atoms:
$$\Omega[D_i]=\{s(X_a^i,S^i_a), s(X_b^i,S^i_b), s(X_c^i,S^i_c)  \}.$$

Let $\Omega=\bigcup_{1\leq i\leq m}\Omega[D_i]$, and denote by $\Omega(D_i)$ the set of all 
atoms of $\Omega$ in which some variable of $D_i$ occurs, i.e,
$$\Omega(D_i)=\{s(X,\alpha)\in\Omega\, |\,  X\in D_i\}.$$

Let $Q$ be the query whose atom-set is 
$\BLOCKS \cup \LINKS \cup \Omega$.

We claim that $Q$ has query width 4 iff $I=(R,\Delta)$ is a positive instance of 
{\tt EXACT COVER BY 3SETS}. 

Let us first prove the {\em if} part. 
Assume that there exist $s$ 3-sets $D^1,\ldots,D^s\in \Delta$ which exactly cover $R$, 
i.e., which form a partition of $R$. We describe a query-decomposition $(T,\lambda)$ 
of $Q$. 

The root $v_{a0}$ of $T$ is labeled by the set of atoms $\BLOCKA^0$.
The root has as unique child a vertex 
$v_{b0}$ labeled by $\BLOCKB^0$.

The decomposition tree is continued as follows. For each $1\leq \ell \leq s$, do the following.

\begin{itemize}
\item Create a vertex $v_{c\ell}$ labeled by 
$\LINK_\ell\cup\Omega[D^\ell]$, and attach $v_{c\ell}$ as a child to 
$v_{b \ell-1}$.

\item For each  remaining atom  $A$ of $\Omega(D^\ell)$, we create a new vertex, label it 
with $\{A\}$, and attach it as a leaf  to $v_{c\ell}$. (Note that these remaining 
atoms, if any, stem from other elements of $\Delta$, given that a variable may 
occur in several 3-sets.) 

\item Then, create a vertex  $v_{a\ell}$ of $T$, label it  by the set of atoms $\BLOCKA^\ell$,
and attach it as  a child of $v_{c \ell}$.
The vertex $V_{a \ell}$, in turn,   has as only child a vertex 
$v_{b\ell}$ labeled by $\BLOCKB^\ell$. 
\end{itemize}
It is not hard to check that $(T,\lambda)$ is indeed a valid query decomposition.

Let us now prove the {\em only-if} part. Assume $(T,\lambda)$ is a 
width 4 query decomposition of the above defined query $Q$.  
By Proposition~\ref{prop:pure-dec}, we also assume, w.l.o.g., that
$(T,\lambda)$ is a pure query decomposition.
Since $Q$ is connected, also $T$ is connected.

We observe a number of relevant facts and make some assumptions.

\begin{description}

\item{FACT 1:} By Lemma~\ref{lem:blocks} for each $0\leq \ell\leq s$, there must exist  
adjacent vertices $v_{a\ell}$ and $v_{b\ell}$ such that 
$\lambda(v_{a\ell})\cup\lambda(v_{b\ell})= \BLOCKA^\ell\cup \BLOCKB^\ell$.

\item{FACT 2:} It holds that 
$S\subseteq var(v_{a\ell})$ and $S\subseteq var(v_{b\ell})$. In fact, if 
this were not the case, then both vertices would miss variables from $S$,
but since all variables of $S$ occur together in other pairs of 
adjacent vertices, this would violate the {\em connectedness condition} 
and is thus impossible.

\item{FACT 3:} From the latter, and from the fact that 
the sets $S'_a,S''_a, S^0_b$, and $S^0_c$ form a partition of $S$,
it follows that each of the vertices 
$v_{a\ell}$ and $v_{b\ell}$ contains a $q$ atom, a $p_a$ atom, a $p_b$ atom,
and a $p_c$ atom. 
Without loss of generality, we  can thus make 
the following assumption.

\item{ASSUMPTION:} For $0\leq \ell\leq s$ we have 
$Z_\ell\in var(v_{a\ell})$ and $Y_\ell\in var(v_{b\ell})$.

\item{FACT 4:} For $1\leq \ell\leq s$, 
there exists a vertex $v_{c\ell}$ 
that lies on the unique path from 
$v_{b\ell-1}$ to $v_{a\ell}$ 
such that 
$\{Y_{\ell -1},Z_\ell\}\subseteq var(v_{c\ell})$.
This can be seen as follows. 
For any variable $\theta$, a {\em $\theta$-path} is a path $\pi$ in 
$T$ such that the variable $\theta$ occurs in the label $\lambda(v)$ 
of any vertex $v$ of $\pi$. 
The atom
$link(Y_{\ell-1},Z_{\ell})$ must belong to the set $\lambda(v'_{c\ell})$
of some vertex $v'_{c\ell}$ of $T$. Clearly, by the 
{\em connectedness} condition,  $v'_{c\ell}$ is connected 
via an $Y_{\ell -1}$-path $\pi_b$ to $v_{b\ell-1}$ and by an
$Z_{\ell}$-path $\pi_a$ to  $v_{a\ell}$. Let $\pi$ denote the 
unique path from $v_{b\ell-1}$ to $v_{a\ell}$. 
Then $\pi$, $\pi_a$, and $\pi_b$ intersect at 
exactly one vertex. This is the desired vertex 
$v_{c\ell}$.

\item{FACT 5:} For $1\leq \ell\leq s$, $S\subseteq var(v_{c\ell})$.
Trivial, because $v_{c\ell}$
lies on a path from 
$v_{b\ell-1}$ to $v_{a\ell}$ and 
$S\subseteq var(v_{b\ell-1})$ and
$S\subseteq var(v_{a\ell})$. The fact follows by the 
{\em connectedness} condition.

\item{FACT 6:}  For $1\leq \ell\leq s$ 
$link(Y_{\ell -1},Z_{\ell})$ belongs to 
$\lambda(v_{c\ell})$ and 
there exists an $i$ with $1\leq i\leq m$ such that 
$\Omega[D_i]\subseteq \lambda(v_{c\ell})$; in summary,
$\lambda(v_{c\ell})=\{link(Y_{\ell -1},Z_{\ell})\}\cup \Omega[D_i]$.
Let us prove this.
By FACT 5 we know that all variables in $S$ must be 
covered by $v_{c\ell}$. However, it also holds that 
$\{Y_{\ell -1},Z_\ell\}\subseteq var(v_{c\ell})$ (see FACT 4).
To cover  the latter variables, there are two alternative choices: 
\begin{enumerate} 
\item both atoms $q(\Pi^{\ell-1}_5,S'_a,Y_{\ell-1})$ and 
$q(\Pi^{\ell}_1,S'_a,Z_{\ell})$ belong to $\lambda(v_{c\ell})$; or
\item the atom $link(Y_{\ell -1},Z_{\ell})$ belongs to 
$\lambda(v_{c\ell})$.
\end{enumerate} 
Choice 1 is impossible: there exist no two other atoms $A,B\in atoms(Q)$
such that $var(A)\cup var(B)\cup S'_a=S$.
We are thus left with Choice 2. Since the atom 
$link(Y_{\ell -1},Z_{\ell})$ does not contain any variable from $S$, 
there must be three other atoms in $\lambda(v_{c\ell})$ that together 
cover $S$. An inspection of the available atoms 
shows that the only possibility of covering $S$ by three atoms is 
via some atom set $\Omega[D_i]$ for $1\leq i\leq m$. The fact is proved.

\item{FACT 7:} 
For $1\leq i<j\leq s$ it holds that 
$v_{ai}$ lies on the unique path in $T$ from 
$v_{ci}$ to $v_{cj}$. 

Consider the edge $\{v_{ai},v_{bi}\}$. If we cut this edge from 
the tree $T$, then we obtain two disconnected trees 
$T_a$ (containing $v_{ai}$) and $T_b$ (containing $v_{bi}$).
Since $v_{ci}$ is connected via a $Z_i$-path to $v_{ai}$, 
but $Z_i$ does not occur in $var(v_{bi})$, it holds that 
$v_{ci}$ is contained in  $T_a$. On the other hand, 
by ``iterative'' application of Fact 4 and of the {\em connectedness
condition} it follows that there is a path $\pi$ from $v_{bi}$ 
to $v_{cj}$ such that for each vertex $v$ of $\pi$ it holds that 
$var(v)\cap Bigvars\neq \emptyset$, where 
$Bigvars=\{Y_h\,|\,i\leq h< j\}\cup \{Z_h\,|\,i< h< j\}$.
Since $var(v_{ai})\cap Bigvars=\emptyset$,
$\pi$ does not traverse $v_{ai}$. It follows that $v_{ci}$ 
belongs to $T_b$. Therefore, the unique path linking 
$v_{ci}$ to $v_{cj}$ goes through the edge
$\{v_{ai},v_{bi}\}$, and thus contains the vertex $v_{ai}$. 

\item{FACT 8:} For $0\leq i< j\leq s$ it holds that 
$var(v_{ci})\cap var(v_{cj})=S$. 
By Fact 7,  $v_{ai}$ lies on the unique path from $v_{ci}$
to $v_{cj}$. Therefore by the {\em connectedness condition}
it holds that $var(v_{ci})\cap var(v_{cj})\subseteq var(v_{ai})$. 
Moreover, by  Fact 6, no variable from $var(v_{ai})-S$ is contained in both 
$var(v_{ci})$ and  $var(v_{cj})$. Thus $var(v_{ci})\cap var(v_{cj})\subseteq S$. 
On the other hand, by Fact 5, $S\subseteq var(v_{ci})$ 
and $S\subseteq var(v_{cj})$, hence, $S\subseteq var(v_{ci})\cap
var(v_{cj})$. 
In summary, we obtain $var(v_{ci})\cap var(v_{cj})=S$. 
\end{description}

For each $1\leq \ell\leq s$, denote by $D^\ell$ the set $D_i$ such that 
$\Omega[D_i]\subseteq \lambda(v_{c\ell})$ (see Fact 6).
By FACT 8 it follows that the sets $D^\ell$ ($1\leq\ell\leq s$)
are mutually disjoint. 
But then the union of these sets is of cardinality $3s=r$, and hence 
the union must coincide with $R$. Thus $s$ subsets out of 
$\Delta$ cover $R$ and $(R,\Delta)$ is a positive instance 
of {\tt EXACT COVER BY 3-SETS}.
\end{proof}

\section{Conjunctive Queries of Bounded Hypertree-Width}
\label{sec:hyper}

\subsection{Hypertree Width}
\label{sec:hw}

Let $Q$ be a (conjunctive) query.
A {\em hypertree for $Q$} is a  triple $\tuple{T,\chi,\lambda}$, where 
$T=(N,E)$ is a rooted tree,
and $\chi$ and $\lambda$ are labeling functions which associate
to each vertex $p\in N$ two sets $\chi(p)\subseteq var(Q)$
and $\lambda(p)\subseteq atoms(Q)$.
If $T'=(N',E')$ is a subtree of $T$, 
we define  $\chi(T')= \bigcup_{v\in N'} \chi(v)$.
We denote the set of vertices $N$ of $T$ by $vertices(T)$, and the
root of $T$ by $root(T)$.
Moreover, for any $p\in N$, $T_p$ denotes the subtree of $T$ rooted at $p$.

\begin{defn}\label{defn:hyper-decomp}\em
A {\em hypertree decomposition} of a conjunctive query $Q$ is 
a hypertree $\tuple{T,\chi,\lambda}$ for $Q$ which satisfies
all the following conditions:
\begin{enumerate}
\item for each atom $A\in atoms(Q)$, there exists $p\in vertices(T)$ 
      such that $var(A)\subseteq \chi(p)$;
\item for each variable $Y\in var(Q)$, the set
 $\{ p\in vertices(T)\; |\;  Y \in \chi(p) \}$
    induces a (connected) subtree of $T$;
\item for each vertex $p\in vertices(T)$, $\chi(p)\subseteq var(\lambda(p))$;
\item for each vertex $p\in vertices(T)$,
$var(\lambda(p)) \cap \chi(T_p) \;\subseteq\; \chi(p)$.
\end{enumerate}

A hypertree decomposition $\tuple{T,\chi,\lambda}$ 
of $Q$ is a {\em complete decomposition} of $Q$ if,
for each atom $A\in atoms(Q)$, there exists $p\in vertices(T)$ 
      such that $var(A)\subseteq \chi(p)$ and $A\in \lambda(p)$.
 
The {\em width} of the hypertree decomposition $\tuple{T,\chi,\lambda}$
is $max_{p\in vertices(T)} |\lambda(p)|$.
The {\em hypertree width} $hw(Q)$ of $Q$ is the minimum width
over all its hypertree decompositions.
\end{defn}
In analogy to join trees and query decompositions, we will refer to
Condition~2 above as the {\em Connectedness Condition}.
Note that, by Condition~1, $\chi(T)=var(Q)$. Hence Condition~4 entails 
that, for $s_0=root(T)$, $var(\lambda(s_0))=\chi(s_0)$.

Intuitively, 
the $\chi$ labeling selects the set of variables to be fixed
in order to split the cycles and achieve acyclicity;
$\lambda(p)$ ``covers'' the variables of $\chi(p)$ by a
set of atoms.
Thus, the relations associated to the atoms of $\lambda(p)$
restrict the range of the variables of $\chi(p)$.
For the evaluation of query $Q$,
each vertex $p$ of the decomposition is replaced by
a new atom whose associated database relation is
the projection on $\chi(p)$ of the join of the relations in $\lambda(p)$.
This way, we obtain a join tree $\JT$ of an acyclic query $Q'$
over database $\DB'$ of size $O(n^k)$,
where $n$ is the input size and $k$ is the width
of the hypertree decomposition.
All the efficient techniques available for acyclic queries can be then
employed for the evaluation of $Q'$.

More technically, Condition 1 and Condition 2 above extend the notion of
tree decomposition \cite{robe-seym-86} from graphs to hypergraphs
(the hypergraph of a query $Q$ groups the variables of the same
atom in one hyperedge \cite{beer-etal-83}).
Thus, the pair $\tuple{T,\chi}$ of a hypertree decomposition
$\tuple{T,\chi,\lambda}$ of a conjunctive query $Q$,
can be seen as the correspondent of a tree decomposition
on the query hypergraph.
However, the treewidth of $\tuple{T,\chi}$ (i.e., the maximum
cardinality of the $\chi$-labels of the vertices of $T$)
is not an appropriate measure of the width of the hypertree decomposition,
because a set of $m$ variables appearing in the same atom
should count $1$ rather than $m$ for the width.
Thus, $\lambda(p)$ provides a set of atoms which
``covers'' $\chi(p)$ and its cardinality gives the measure
of the width of vertex $p$.
It is worthwhile noting that $\tuple{T,\lambda}$ may violate
the classical connectedness condition usually imposed on the variables
of the join trees,
as it is allowed that a variable $X$ appears in both
$\lambda(p)$ and $\lambda(q)$ while it does not appear
in $\lambda(s)$, for some vertex $s$ on the path from $p$ to $q$ in $T$.
However, this violation is not a problem, as
the variables in $var(\lambda(p))-\chi(p)$
are meaningless and can be projected out before starting
the query evaluation process, because the role of $\lambda(p)$
is just that of providing a binding for the variables of $\chi(p)$.

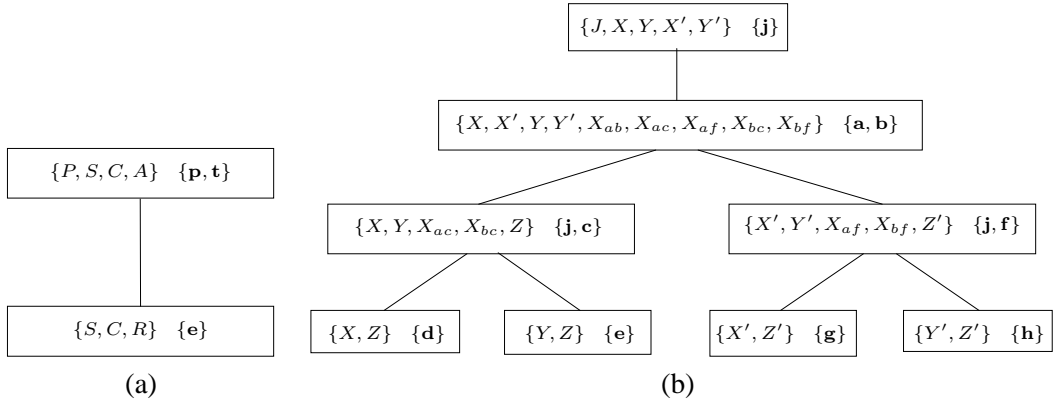
\begin{figure}[h]
\center
\setlength{\unitlength}{0.00029167in}
\begingroup\makeatletter\ifx\SetFigFont\undefined
\def\x#1#2#3#4#5#6#7\relax{\def\x{#1#2#3#4#5#6}}%
\expandafter\x\fmtname xxxxxx\relax \def\y{splain}%
\ifx\x\y   
\gdef\SetFigFont#1#2#3{%
  \ifnum #1<17\tiny\else \ifnum #1<20\small\else
  \ifnum #1<24\normalsize\else \ifnum #1<29\large\else
  \ifnum #1<34\Large\else \ifnum #1<41\LARGE\else
     \huge\fi\fi\fi\fi\fi\fi
  \csname #3\endcsname}%
\else
\gdef\SetFigFont#1#2#3{\begingroup
  \count@#1\relax \ifnum 25<\count@\count@25\fi
  \def\x{\endgroup\@setsize\SetFigFont{#2pt}}%
  \expandafter\x
    \csname \romannumeral\the\count@ pt\expandafter\endcsname
    \csname @\romannumeral\the\count@ pt\endcsname
  \csname #3\endcsname}%
\fi
\fi\endgroup
{\renewcommand{\dashlinestretch}{30}
\begin{picture}(4824,3789)(0,-10)
\path(12,912)(4812,912)(4812,12)
	(12,12)(12,912)
\path(12,3762)(4812,3762)(4812,2862)
	(12,2862)(12,3762)
\path(2412,912)(2404,2845)
\put(2412,3237){\makebox(0,0)[b]{\smash{{{\SetFigFont{7}{8.4}{rm}$\{P,S,C,A\}\;\;\;\bf\{p, t\}$}}}}}
\put(2412,387){\makebox(0,0)[b]{\smash{{{\SetFigFont{7}{8.4}{rm}$\{S,C,R\}\;\;\;\bf\{e\}$}}}}}
\end{picture}
}
\setlength{\unitlength}{0.00027500in}
\begingroup\makeatletter\ifx\SetFigFont\undefined
\def\x#1#2#3#4#5#6#7\relax{\def\x{#1#2#3#4#5#6}}%
\expandafter\x\fmtname xxxxxx\relax \def\y{splain}%
\ifx\x\y   
\gdef\SetFigFont#1#2#3{%
  \ifnum #1<17\tiny\else \ifnum #1<20\small\else
  \ifnum #1<24\normalsize\else \ifnum #1<29\large\else
  \ifnum #1<34\Large\else \ifnum #1<41\LARGE\else
     \huge\fi\fi\fi\fi\fi\fi
  \csname #3\endcsname}%
\else
\gdef\SetFigFont#1#2#3{\begingroup
  \count@#1\relax \ifnum 25<\count@\count@25\fi
  \def\x{\endgroup\@setsize\SetFigFont{#2pt}}%
  \expandafter\x
    \csname \romannumeral\the\count@ pt\expandafter\endcsname
    \csname @\romannumeral\the\count@ pt\endcsname
  \csname #3\endcsname}%
\fi
\fi\endgroup
{\renewcommand{\dashlinestretch}{30}
\begin{picture}(14154,6762)(0,-10)
\path(6987,5880)(6987,4890)
\path(12747,885)(11172,1965)
\path(4917,6735)(9102,6735)(9102,5835)
	(4917,5835)(4917,6735)
\path(10902,2910)(7392,3945)
\path(7617,885)(10407,885)(10407,12)
	(7617,12)(7617,885)
\path(11307,885)(14142,885)(14142,102)
	(11307,102)(11307,885)
\path(8967,885)(10542,1965)
\path(3207,2910)(6582,3945)
\path(5142,885)(3577,1982)
\path(1407,885)(2982,1965)
\path(3702,885)(6492,885)(6492,45)
	(3702,45)(3702,885)
\path(12,885)(2847,885)(2847,75)
	(12,75)(12,885)
\path(327,2910)(6177,2910)(6177,1965)
	(327,1965)(327,2910)
\path(7977,2910)(13827,2910)(13827,1965)
	(7977,1965)(7977,2910)
\path(2442,4890)(11667,4890)(11667,3945)
	(2442,3945)(2442,4890)
\put(6987,4350){\makebox(0,0)[b]{\smash{{{\SetFigFont{7}{8.4}{rm}$\{X,X',Y,Y',X_{ab},X_{ac},X_{af},X_{bc},X_{bf}\}\;\;\;\bf\{a, b\}$}}}}}
\put(10902,2415){\makebox(0,0)[b]{\smash{{{\SetFigFont{7}{8.4}{rm}$\{X',Y',X_{af},X_{bf},Z'\}\;\;\;\bf\{j, f\}$}}}}}
\put(12747,345){\makebox(0,0)[b]{\smash{{{\SetFigFont{7}{8.4}{rm}$\{Y',Z'\}\;\;\;\bf\{h\}$}}}}}
\put(8967,345){\makebox(0,0)[b]{\smash{{{\SetFigFont{7}{8.4}{rm}$\{X',Z'\}\;\;\;\bf\{g\}$}}}}}
\put(3207,2370){\makebox(0,0)[b]{\smash{{{\SetFigFont{7}{8.4}{rm}$\{X,Y,X_{ac},X_{bc},Z\}\;\;\;\bf\{j, c\}$}}}}}
\put(5142,345){\makebox(0,0)[b]{\smash{{{\SetFigFont{7}{8.4}{rm}$\{Y,Z\}\;\;\;\bf\{e\}$}}}}}
\put(1452,345){\makebox(0,0)[b]{\smash{{{\SetFigFont{7}{8.4}{rm}$\{X,Z\}\;\;\;\bf\{d\}$}}}}}
\put(7032,6195){\makebox(0,0)[b]{\smash{{{\SetFigFont{7}{8.4}{rm}$\{J,X,Y,X',Y'\}\;\;\;\bf\{j\}$}}}}}
\end{picture}
}\\
\mbox{\hspace{3cm}}\mbox{(a)}\mbox{\hspace{6.7cm}}\mbox{(b)}
\hfill\mbox{}
\caption{A 2-width hypertree decomposition of (a) query $Q_1$; and
  (b) query $Q_5$}
\label{fig:HDvsQD}
\end{figure}

\begin{example}\label{ex:decomposition} \em
The hypertree width of the cyclic query $Q_1$ of Example~\ref{ex:queries}
is 2; a (complete) 2-width hypertree decomposition of $Q_1$ is shown in
Figure~\ref{fig:HDvsQD}.a.

Consider the following conjunctive query $Q_5$:
\[
\begin{array}{ll}
 ans \leftarrow & a(X_{ab},X,X',X_{ac},X_{af})\wedge 
  b(X_{ab},Y,Y',X_{bc},X_{bf})\wedge 
  c(X_{ac},X_{bc},Z)\wedge  d(X,Z)\wedge e(Y,Z)\wedge \\
  & \wedge f(X_{af},X_{bf},Z')\wedge
  g(X',Z')\wedge h(Y',Z')\wedge j(J,X,Y,X',Y')
\end{array}
\]

$Q_5$ is clearly cyclic, and thus $hw(Q_5)>1$ (as only acyclic queries
have hypertree width equals 1).
Figure~\ref{fig:HDvsQD} shows a (complete) hypertree decomposition of $Q_5$
having width 2, hence $hw(Q_5)=2$.
\end{example}

Definition \ref{defn:hyper-decomp} does not require
the presence of \underline{all} query atoms in a decomposition $\HD$,
as it is sufficient that every atom is "covered" by some vertex $p$ of $\HD$
(i.e., its variables are included in $\chi(p)$).
However, every missing atom can be easily added to complete
decompositions.

\begin{lem}\label{lem:complete}
Given a conjunctive query $Q$,
every $k$-width hypertree decomposition of $Q$
can be transformed in Logspace into 
a $k$-width complete hypertree decomposition of $Q$.
\end{lem}
\begin{proof} 
Let $Q$ be a conjunctive query and $\HD=\tuple{T,\chi,\lambda}$ a
hypertree decomposition of $Q$. 
In order to transform $\HD$ into a complete decomposition,
modify $\HD$ as follows.
For each atom $A\in atoms(Q)$ such that  
no vertex $q\in vertices(T)$ satisfies 
$var(A)\subseteq \chi(q)$ and $A\in\lambda(q)$, create a new
vertex $v_A$ with $\lambda(v_A):=\{A\}$ and $\chi(v_A)=var(A)$, 
and attach $v_A$ as a new
child of a vertex $p\in vertex(T)$ s.t. $var(A)\subseteq \chi(p)$.
(By Condition~1 of Definition~\ref{defn:hyper-decomp} such a $p$ must exist.)

This transformation is obviously feasible in Logspace.
\end{proof}

The acyclic queries are precisely the queries of hypertree width one.

\begin{theo}\label{th:acyclic}
A conjunctive query $Q$ is acyclic if and only if
$hw(Q)=1$.
\end{theo}
\begin{proof}
({\em Only if part.}) If $Q$ is an acyclic query, there exists a join
tree $\JT(Q)$ for $Q$. Let $T$ be a tree, and 
$f$ a bijection from $vertices(\JT(Q))$ to $vertices(T)$ such that,
for any $p,q\in vertices(\JT(Q))$, there is an edge between $p$ and
$q$ in $\JT(Q)$ if and only if there is an edge between $f(p)$ and $f(q)$
in $T$. Moreover, let $\lambda$ be the following labeling function:
If $p$ is a vertex of $\JT(Q)$ and $A$ is the atom
of $Q$ associated to $p$, then $\lambda(f(p))=\{A\}$.
For any vertex $p'\in vertices(T)$ define $\chi(p')=var(\lambda(p'))$.
Then, $\tuple{T,\chi,\lambda}$ is clearly a width 1 
hypertree-decomposition of $Q$.

({\em If part.}) Let $\HD=\tuple{T,\chi,\lambda}$ be a width 1 
hypertree-decomposition of $Q$. W.l.o.g., assume that $\HD$ is a
complete hypertree decomposition. Since $\HD$ has width 1, all the
$\lambda$ labels are singletons, i.e., $\lambda$ associate one atom of
$Q$ to each vertex of $T$. 

We next show how to trasform $\HD$ into a width 1 complete hypertree
decomposition of $Q$ such that, for any vertex $p\in vertices(T)$,
$\chi(p)=var(A)$, where $\{A\}=\lambda(p)$, and $p$ is the unique
vertex labeled with the atom $A$.

Choose any total ordering $\prec$ of the vertices of $T$.
For any atom $A\in atoms(Q)$, denote by $v(A)$ the $\prec$-least vertex of $T$ 
such that $\chi(v(A))=var(A)$ and $\lambda(v(A))=\{A\}$.
The existence of such a vertex is guaranteed by definition of
complete hypertree decomposition and by the hypothesis that
every $\lambda$ label consists of exactly one atom.

For any atom $A\in atoms(Q)$, 
and for any vertex $p\neq v(A)$ such that $\lambda(p)=\{A\}$, 
perform the following actions.
For any child $p'$ of $p$, delete the edge between $p$ and $p'$ and
let $p'$ be a new child of $v(A)$, hence the subtree $T_{p'}$ is now
attached to $v(A)$. Then, delete vertex $p$.
By Condition~3 of Definition~\ref{defn:hyper-decomp},
$\chi(p)\subseteq var(\lambda(p))$. Since
$var(\lambda(p))=var(A)=\chi(v(A))$, we get $\chi(p)\subseteq \chi(v(A))$.
Then, it is easy to see that the (transformed) tree $T$ satisfies
the connectedness condition.

Eventually, we obtain a new hypertree $H'=\tuple{T',\chi,\lambda}$ 
such that $vertices(T')\subseteq vertices(T)$ and $H'$ has the
following properties:
(i) for any $A\in atoms(Q)$, there exists exactly one vertex $p=v(A)$ of $T'$
such that $\lambda(p)=\{A\}$ and $\chi(p)=var(A)$; (ii)
for any vertex $p$ of $T'$, $p=v(A)$ holds, for some
$A\in atoms(Q)$;
(iii) $H'$ satisfies the connectedness condition.
Thus, $H'$ clearly corresponds to a join tree of $Q$.
\end{proof}

\subsection{Efficient Query Evaluation}

\begin{lem}\label{lem:HDvsJT}
Let $Q$ be a Boolean conjunctive query over a database $\DB$,
and $\HD=\tuple{T,\chi,\lambda}$ a hypertree decomposition
of $Q$ of width $k$. Then, there exists $Q',\DB',\JT$ such that:
\begin{enumerate}
\item
$Q'$ is an acyclic (Boolean) conjunctive query answering 'yes' on
database $\DB'$ iff the answer of $Q$ on $\DB$ is 'yes'.
\item
$||\tuple{Q',\DB',\JT}||= O(||\tuple{Q,\DB,\HD}||^k)$.
\item
$\JT$ is a join tree of the query $Q'$.
\item
$\tuple{Q',\DB',\JT}$ is logspace computable from
$\tuple{Q,\DB,\HD}$.
\end{enumerate}
\end{lem}
\begin{proof} 
Let $Q$ be a Boolean conjunctive query over a database $\DB$,
and $\HD=\tuple{T,\chi,\lambda}$ a hypertree decomposition
of $Q$ of width $k$. 
From Lemma \ref{lem:complete}, we can assume that $\HD=\tuple{T,\chi,\lambda}$
is a complete decomposition of $Q$.
W.l.o.g., we also assume $Q$ does not contain any atom $A$ such that
$var(A)=\emptyset$. 

Note that $Q$ evaluates to {\em true} on $\DB$ if and only if 
$\Join_{A\in atoms(Q)} rel(A)$ is a non-empty relation, where 
$rel(A)$ denotes the relation of $\DB$ associated to the atom $A$,
and $\Join$ is the natural join operation 
(with common variables acting as join attributes).

For each vertex $p\in vertices(T)$ define a query $Q(p)$ and a database 
$\DB(p)$ as follows. For each atom $A\in \lambda(p)$:
\begin{itemize}
\item If $var(A)\subseteq\chi(p)$, then $A$ occurs in $Q(p)$ and
 $rel(A)$ belongs to $\DB(p)$;
\item if $(var(A)\cap\chi(p))\neq\emptyset$, then $Q(p)$ 
  contains a new atom $A'$ such that $var(A')=var(A)\cap\chi(p)$, 
  and $\DB(p)$ contains the corresponding relation $rel(A')$, which is 
  the projection of $rel(A)$ on the set of attributes corresponding to the 
  variables in $var(A')$;
\end{itemize}

Now, consider the following query $\bar Q$ on the database 
$\bar{\DB} = \bigcup_{p\in vertices(T)} \DB(p)$. 
\[ \bar Q:\quad \bigwedge_{p\in vertices(T)} Q(p) \]

By the associative and commutative properties of natural joins,
and by the fact that $\HD$ is a complete hypertree decomposition,
it immediately follows that 
$\bar Q$ on $\bar{\DB}$ is equivalent to $Q$ on $\DB$.

We build $\tuple{Q',\DB',\JT}$ as follows.
$\JT$ has exactly the same tree shape of $T$.
For each vertex $p$, there is precisely one vertex $p'$ in $\JT$,
and one relation $P'$ in $\DB'$.
$p'$ is an atom having $\chi(p)$ as arguments and its corresponding relation
$P'$ in $\DB'$ is the result of the query $Q(p)$ on $\DB(p)$.
$Q'$ is the conjunction of all vertices (atoms) of $JT$.
$Q'$ on $\DB'$ is clearly equivalent to $Q$ on $\DB$, and
$\JT$ is a join tree of $Q'$.
Moreover,
$||\DB'|| = O(||\DB||^k)$, $||JT||= O(||\HD||)$ and $||Q'||= O(||\HD||)$;
thus, $||\tuple{Q',\DB',\JT}||= O(||\tuple{Q,\DB,\HD}||^k)$.

The transformation is clearly feasible in Logspace.
\end{proof}

\begin{theo}\label{th:hdcfl}
Given a database $\DB$, a Boolean conjunctive query
$Q$, and a k-width hypertree decomposition of $Q$ for a fixed
constant $k>0$, deciding whether $Q$ evaluates to {\em true} on $\DB$
is $\LCFL$-complete.
\end{theo}

\begin{theo}\label{th:hdpoly}
Given a database $\DB$, a (non-Boolean) conjunctive query
$Q$, and a k-width hypertree decomposition of $Q$ for a fixed
constant $k>0$, the answer of $Q$ on $\DB$
can be {\em computed} in time polynomial in the combined size of
the input instance and of the output relation.
\end{theo}

\noindent
{\bf Remark.} In this section we demonstrated
that k-bounded hypertree-width queries
are efficiently computable, once a k-width hypertree decomposition
of the query is given as (additional) input.
In Section~\ref{sec:main}, we will strenghten these results
showing that providing the hypertree decomposition in input is unnecessary,
as, different from query decompositions,
a hypertree decomposition can be computed very efficiently
(in $\FCFL$, i.e., in functional $\LCFL$).

\section{Bounded Hypertree Decompositions are Efficiently Computable}

\subsection{Normal form}

Let $V\subseteq var(Q)$ be a set of variables, and $X,Y\in var(Q)$ 
a pair of  variables occurring in $Q$,
then $X$ is \adj{V}\  to $Y$ if there exists an atom $A\in atoms(Q)$ such
that $\{X,Y\}\subseteq (var(A)- V)$.
A \spath{V}\ 
$\pi$ from $X$ to $Y$ consists of a sequence $X=X_0,\ldots,X_h=Y$
of variables and a sequence of atoms 
$A_0,\ldots,A_{h-1}$ ($h\geq 0$)  such that: $X_{i}$ is \adj{V}\ to
$X_{i+1}$ and
$\{X_i,X_{i+1}\}\subseteq var(A_i)$, for each $i\in [0...h\mbox{-}1]$.
We denote by $var(\pi)$ (resp. $atoms(\pi)$) 
the set of variables (atoms) occurring in the sequence
$X_0,\ldots,X_h$ ($A_0,\ldots,A_{h-1}$).

Let $V\subseteq var(Q)$ be a set of variables occurring in a query $Q$.
A set $W\subseteq var(Q)$ 
of variables is \connected{V}\ if $\forall X,Y\in W$ there
is a \spath{V}\ from $X$ to $Y$.
A {\em \component{V}} is a maximal \connected{V}\ 
non-empty set of variables $W\subseteq (var(Q)-V)$.

Note that the variables in $V$ do not belong to any \component{V}\ 
(i.e., $V \cap C =\emptyset$ for each \component{V}\ $C$).

Let $C$ be a \component{V}\ for some set of variables $V$. 
We define:
\[
\begin{array}{lll}
atoms(C) & := & \{ A\in atoms(Q) \;|\; var(A)\cap C\neq\emptyset\}.
\end{array}
\]

Note that, for any set of variables $V$,
and for every atom $A\in atoms(Q)$ such that $var(A)\not\subseteq V$,
there exists exactly one \component{V}\ $C$ of $Q$ such that
$A\in atoms(C)$.

Furthermore, let $H=\tuple{T,\chi,\lambda}$ be a hypertree of $Q$
and $V\subseteq var(Q)$ a set of variables.
We define 
$vertices(V,H)=\{ p\in vertices(T) \;|\; \chi(p) \cap V \neq \emptyset\}$.

For any vertex $v$ of $T$, we will often use $v$ as a synonym of
$\chi(v)$. In particular, $\component{v}$ denotes
$\component{\chi(v)}$; the term \spath{v}\ is a synonym of
\spath{\chi(v)}; and so on.

\begin{defn}\label{defn:nf}
{\em
A hypertree decomposition $\HD=\tuple{T,\chi,\lambda}$
of a conjunctive query $Q$ is
in {\em normal form} ({\em NF}) if 
for each vertex $r\in vertices(T)$, and for each child $s$ of $r$,
all the following conditions hold:
\begin{enumerate}
\item
there is (exactly) one $\component{r}$ $C_r$ such that
$\chi(T_s)\;=\; C_r\cup (\chi(s)\cap \chi(r))$;
\item
$\chi(s)\cap C_r \neq \emptyset$, where $C_r$ is the \component{r} 
satisfying Point 1;
\item $var(\lambda(s))\cap \chi(r) \; \subseteq \; \chi(s)$.
\end{enumerate}
}
\end{defn}

Note that Condition~2 above entails that, 
for each vertex $r\in vertices(T)$, and for each child $s$ of $r$,
$\chi(s)\not\subseteq \chi(r)$. Indeed, $C_r\cap \chi(r)=\emptyset$,
and $s$ must contain some variable belonging to the \component{r}\ $C_r$.

\begin{lem}\label{lem:path}
Let $\HD=\tuple{T,\chi,\lambda}$ be a hypertree decomposition of 
a conjunctive query $Q$.
Let $r$ be a vertex of\/ $T$, let $s$ be a child of $r$, and let
$C$ be an \component{r}\ of $Q$ such that $C\cap \chi(T_s)\neq\emptyset$.
Then, $vertices(C,\HD)\subseteq vertices(T_s)$.
\end{lem}
\begin{proof}
For any subtree $T'$ of $T$, let $covered(T')$ denote the set
$\{ A\in atoms(Q) \;|\; var(A)\subseteq\chi(v)$ 
                                 for some $v\in vertices(T')\}$.

Since $C\cap \chi(T_s)\neq\emptyset$, 
there exists a vertex $p\in vertices(T_s)$ 
which also belongs to $vertices(C,\HD)$.
We proceed by contradiction. Assume there exists some vertex 
$q\in vertices(C,\HD)$ such that $q\not\in vertices(T_s)$.
By definition of $vertices(C,\HD)$, there exists a pair of variables
$\{X,Y\} \subseteq C$ such that $X\in \chi(p)$ and $Y\in \chi(q)$.
Since $X,Y \in C$, there exists an $\spath{r}$ $\pi$ from $X$ to $Y$
consisting of a sequence of variables
$X=X_0,\ldots,X_i,X_{i+1},\ldots,X_\ell=Y$, and a sequence of atoms
$A_0,\ldots,A_i,A_{i+1},\ldots,A_{\ell -1}$.

Note that $Y\not\in \chi(T_s)$. Indeed, $Y\not\in \chi(r)$, hence any
occurrence of $Y$ in $\chi(v)$, for some vertex $v$ of $T_s$, would violate
Condition~2 of Definition~\ref{defn:hyper-decomp}. 
Similarly, $X$ only occurs as a variable in $\chi(T_s)$.
As a consequence, 
$A_0\in covered(T_s)$ (by Condition~1 of Definition~\ref{defn:hyper-decomp}) 
and $A_{\ell -1}\not\in covered(T_s)$,
hence the \spath{r}\ $\pi$ leaves $T_s$, i.e., 
$atoms(\pi)\not\subseteq covered(T_s)$. 

Assume w.l.o.g. that the atoms $A_i$, $A_{i+1} \in atoms(\pi)$ 
form the ``frontier''
of this path w.r.t. $T_s$, i.e., $A_i\in covered(T_s)$ and
$A_{i+1}\not\in covered(T_s)$, and consider the variable $X_{i+1}$, which
occurs in both $A_i$ and $A_{i+1}$.  $X_{i+1}$ belongs to $C$, hence
it does not occur in $\chi(r)$, and this immediately yields 
a contradiction to 
Condition~2 of Definition~\ref{defn:hyper-decomp}.
\end{proof}

\begin{lem}\label{lem:connected}
Let $\HD=\tuple{T,\chi,\lambda}$ be a hypertree decomposition of 
a conjunctive query $Q$ and $r\in vertices(T)$. 
If\/ $V$ is an \connected{r}\ set of variables
in $var(Q)-\chi(r)$, then $vertices(V,\HD)$ induces a (connected) subtree of
$T$.
\end{lem}
\begin{proof}
We use induction on $|V|$. 

{\em Basis.} If $|V|=1$, then $V$ is a singleton, and the statement
follows from Condition~2 of Definition~\ref{defn:hyper-decomp}.

{\em Induction Step.} Assume the statement is established for
set of variables having 
cardinalities $c\leq h$. Let $V$ be an \connected{r}\ set of variables
such that $|V|= h+1$, and let $X\in V$ be any variable of $V$ such
that $V-\{X\}$ remains \connected{r}. (It is easy to see that such a
variable  exists.)
By the induction hypothesis, $vertices(V-\{X\},\HD)$ induces a connected
subtree of $T$. Moreover, $\{X\}$ is a singleton,
thus $vertices(\{X\},\HD)$ induces a connected subtree of $T$, too.

Since $X\in V$, $V$ is $\connected{r}$, and $|V|>1$,
there exists a variable $Y\in V-\{X\}$ which is $\adj{r}$ to $X$.
Hence, there exists an atom $A\in atoms(Q)$
such that $\{X,Y\}\subseteq var(A)$.
By Condition~1 of Definition~\ref{defn:hyper-decomp}, there exists a
vertex $p\in vertices(T)$ such that $var(A)\subseteq \chi(p)$.
Note that 
$vertices(V,\HD)= vertices(V-\{X\},\HD)\cup vertices(\{X\},\HD)$, and
$p$ belongs to both $vertices(V-\{X\},\HD)$ and 
$vertices(\{X\},\HD)$.  
Then, both sets induce connected subgraphs of $T$ that are, moreover,
connected to each other via the vertex $p$.
Thus,
$vertices(V,\HD)$ induces a connected subgraph of $T$, and hence a
subtree, because $T$ is a tree.
\end{proof}

\begin{figure}[h]
\center
\setlength{\unitlength}{0.00041667in}
\begingroup\makeatletter\ifx\SetFigFont\undefined
\def\x#1#2#3#4#5#6#7\relax{\def\x{#1#2#3#4#5#6}}%
\expandafter\x\fmtname xxxxxx\relax \def\y{splain}%
\ifx\x\y   
\gdef\SetFigFont#1#2#3{%
  \ifnum #1<17\tiny\else \ifnum #1<20\small\else
  \ifnum #1<24\normalsize\else \ifnum #1<29\large\else
  \ifnum #1<34\Large\else \ifnum #1<41\LARGE\else
     \huge\fi\fi\fi\fi\fi\fi
  \csname #3\endcsname}%
\else
\gdef\SetFigFont#1#2#3{\begingroup
  \count@#1\relax \ifnum 25<\count@\count@25\fi
  \def\x{\endgroup\@setsize\SetFigFont{#2pt}}%
  \expandafter\x
    \csname \romannumeral\the\count@ pt\expandafter\endcsname
    \csname @\romannumeral\the\count@ pt\endcsname
  \csname #3\endcsname}%
\fi
\fi\endgroup
{\renewcommand{\dashlinestretch}{30}
\begin{picture}(14124,6114)(0,-10)
\path(2112,6087)(1212,4362)(3012,4362)(2112,6087)
\path(1512,4362)(1512,2637)(12,2637)
	(387,3087)(1512,4362)
\path(2712,3837)(4212,3837)(4212,3087)
	(2712,3087)(2712,3837)
\path(4062,2487)(5562,2487)(5562,1737)
	(4062,1737)(4062,2487)
\path(3012,3087)(3012,1362)(1512,1362)
	(1887,1812)(3012,3087)
\path(4812,1737)(3912,12)(5712,12)(4812,1737)
\path(2562,4362)(3462,3837)
\path(3912,3087)(4812,2487)
\path(10812,6087)(9912,4362)(11712,4362)(10812,6087)
\path(10212,4362)(10212,2637)(8712,2637)
	(9087,3087)(10212,4362)
\path(11262,4362)(12162,3837)
\thicklines
\path(5562,3837)(8637,3837)
\blacken\path(8412.000,3762.000)(8637.000,3837.000)(8412.000,3912.000)(8479.500,3837.000)(8412.000,3762.000)
\thinlines
\path(11712,3087)(11712,1362)(10212,1362)
	(10587,1812)(11712,3087)
\path(12612,3087)(12612,1362)(14112,1362)
	(13737,1812)(12612,3087)
\path(11262,3837)(13062,3837)(13062,3087)
	(11262,3087)(11262,3837)
\put(12162,3387){\makebox(0,0)[b]{\smash{{{\SetFigFont{8}{9.6}{rm}$r$}}}}}
\put(10812,4962){\makebox(0,0)[b]{\smash{{{\SetFigFont{8}{9.6}{rm}$\alpha$}}}}}
\put(9687,3087){\makebox(0,0)[b]{\smash{{{\SetFigFont{8}{9.6}{rm}$\beta$}}}}}
\put(987,3087){\makebox(0,0)[b]{\smash{{{\SetFigFont{8}{9.6}{rm}$\beta$}}}}}
\put(2487,1737){\makebox(0,0)[b]{\smash{{{\SetFigFont{8}{9.6}{rm}$\gamma$}}}}}
\put(4812,462){\makebox(0,0)[b]{\smash{{{\SetFigFont{8}{9.6}{rm}$\delta$}}}}}
\put(3462,3387){\makebox(0,0)[b]{\smash{{{\SetFigFont{8}{9.6}{rm}$r$}}}}}
\put(4812,2037){\makebox(0,0)[b]{\smash{{{\SetFigFont{8}{9.6}{rm}$s$}}}}}
\put(6837,4062){\makebox(0,0)[b]{\smash{{{\SetFigFont{8}{9.6}{rm}If $\chi(s)\subseteq \chi(r)$}}}}}
\put(2112,4962){\makebox(0,0)[b]{\smash{{{\SetFigFont{8}{9.6}{rm}$\alpha$}}}}}
\put(11187,1737){\makebox(0,0)[b]{\smash{{{\SetFigFont{8}{9.6}{rm}$\gamma$}}}}}
\put(13137,1737){\makebox(0,0)[b]{\smash{{{\SetFigFont{8}{9.6}{rm}$\delta$}}}}}
\end{picture}
}
\caption{Normalizing a hypertree decomposition}\label{fig:nf}
\end{figure}
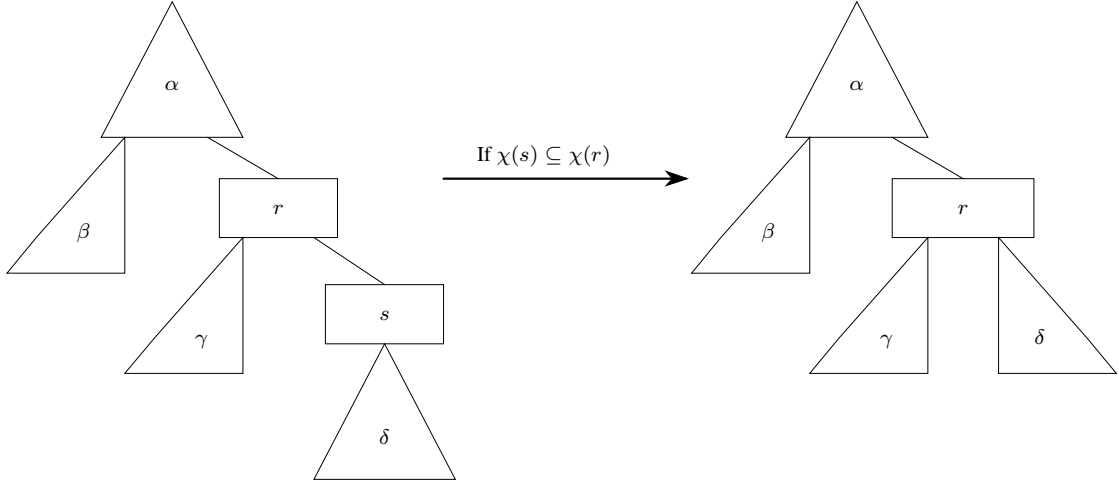

\begin{theo}\label{theo:nf}
For each $k$-width hypertree decomposition of a conjunctive query $Q$ 
there exists a $k$-width hypertree decomposition of $Q$ in
normal form.
\end{theo}
\begin{proof} 
Let $\HD=\tuple{T,\chi,\lambda}$ be any $k$-width hypertree 
decomposition of $Q$.
We show how to transform $\HD$ into a $k$-width hypertree
decomposition in normal form.

Assume 
there exist two vertices $r$ and $s$ s.t. $s$ is a child of $r$, 
and $s$ violates any condition of Definition~\ref{defn:nf}.
If $s$ satisfies Condition~1, but violates Condition~2, then 
$\chi(s)\subseteq \chi(r)$ holds.
In this case, simply eliminate
vertex $s$ from the tree as shown in Figure~\ref{fig:nf}.
It is immediate to see that this transformation is correct.

Assume $T_s$ does not meet Condition~1 of Definition~\ref{defn:nf},
and let $C_1,\ldots,C_h$ be all the \component{r}s
containing some variable occurring in $\chi(T_s)$. Hence, 
$\chi(T_s)\subseteq (\bigcup_{1\leq i\leq h} C_i\; \cup\; \chi(r))$.
For each \component{r}\ $C_i$ $(1\leq i\leq h)$, consider the set of
vertices $vertices(C_i,\HD)$.
Note that, by Lemma~\ref{lem:connected}, 
$vertices(C_i,\HD)$ induces a subtree of $T$, and 
by Lemma~\ref{lem:path}, $vertices(C_i,\HD)\subseteq vertices(T_s)$,
hence $vertices(C_i,\HD)$ induces in fact a subtree of $T_s$.

For each vertex $v\in vertices(C_i,\HD)$
define a new vertex $new(v,C_i)$, and let 
$\lambda(new(v,C_i))= \lambda(v)$ and
$\chi(new(v,C_i))= \chi(v) \cap ( C_i \cup \chi(r) )$.
Note that $\chi(new(v,C_i))\neq\emptyset$, because 
by definition of $vertices(C_i,\HD)$, $\chi(v)$ contains some variable
belonging to $C_i$.
Let $N_i=\{ new(v,C_i) \;|\; v\in vertices(C_i,\HD)\}$.
Moreover, 
for any $C_i$ $(1\leq i\leq h)$, let $T_i$ denote the (directed) graph
$(N_i,E_i)$ such that $new(p,C_i)$ is a child of $new(q,C_i)$ 
iff $p$ is is a child of $q$ in $T$.
$T_i$ is clearly isomorphic to the subtree of $T_s$ induced by
$vertices(C_i,\HD)$, hence $T_i$ is a tree, as well. 

Now, transform the hypertree decomposition $\HD$ as follows.
Delete every vertex in $vertices(T_s)$ from $T$, 
and attach to $r$ every tree $T_i$ for $1\leq i\leq h$.
Intuitively, we replace the subtree $T_s$ by the set of trees
$\{T_1,\ldots,T_h\}$.
By construction, $T_i$ contains a vertex $new(v,C_i)$ 
for each vertex $v$ belonging to $vertices(C_i,\HD)$
($1\leq i\leq h$). Then, if we let $children(r)$ denote the set
of children of $r$ in the new tree $T$ obtained after the
transformation above,
it holds that for any $s'\in children(r)$, there exists an
\component{r}\ $C$ of $Q$ such that $vertices(T_{s'})=vertices(C,\HD)$,
and $\chi(T_{s'})\subseteq (C\cup \chi(r))$.
Furthermore, it is easy to verify that all the conditions of
Definition~\ref{defn:hyper-decomp} are preserved during this
transformation. As a consequence, Condition~2 of 
Definition~\ref{defn:hyper-decomp} immediately entails that
$(\chi(T_{s'}) \cap \chi(r)) \subseteq \chi(s')$.
Hence, $\chi(T_{s'}) = C \cup (\chi(s')\cap \chi(r))$.
Thus, any child of $r$ satisfies both Condition 1 and Condition
2 of Definition~\ref{defn:nf}.

Now, assume that some vertex $v\in children(r)$ violates Condition~3
of Definition~\ref{defn:nf}. Then, add to the label
$\chi(v)$ the set of variables $var(\lambda(v))\cap \chi(r)$.
Because variables in $\chi(r)$ induce connected subtrees of $T$, and
$\chi(r)$ does not contain any variable occurring in some
\component{r}, this further transformation never invalidates any other
condition.
Moreover, no new vertex is labeled by a set of atoms
with cardinality greater than $k$, then 
we get in fact a legal $k$-width hypertree decomposition.

Clearly, $root(T)$ cannot violate any of the normal form conditions,
because it has no parent in $T$.
Moreover, the transformations above never change the parent $r$ of a
violating vertex $s$. 
Thus, if we apply such a transformation to the children of $root(T)$,
and iterate the process on the new children of $root(T)$, and so on,
we eventually gets
a new $k$-width hypertree decomposition $\tuple{T',\lambda'}$ of $Q$
in normal form.
\end{proof}

If $\HD=\tuple{T,\chi,\lambda}$ is an NF hypertree decomposition of 
a conjunctive query $Q$,
we can associate a set $treecomp(s)\subseteq var(Q)$ to each
vertex $s$ of $T$ as follows.
\begin{itemize}
\item If $s=root(T)$, then $treecomp(s)=var(Q)$;
\item otherwise, let $r$ be the father of $s$ in $T$; then,
$treecomp(s)$ is the (unique) \component{r}\ $C$ such that 
$\chi(T_s)\;=\; C \cup (\chi(s)\cap \chi(r))$.
\end{itemize}
Note that, since $s\in vertices(T_s)$, also
$\chi(T_s)\;=\; C \cup \chi(s)$ holds.

\begin{lem}\label{lem:partition}
Let $\HD=\tuple{T,\chi,\lambda}$ be an NF hypertree decomposition of 
a conjunctive query $Q$, $v$ a vertex of $T$, and $W= treecomp(v)-\chi(v)$.
Then, for any \component{v}\ $C$ such that $(C\cap W)\neq\emptyset$,
$C\subseteq W$ holds. 

Therefore,
the set ${\cal C}=\{ C'\subseteq var(Q)\;|\;$ $C'$ is a \component{v}\ 
and $C'\subseteq treecomp(v)\}$ is a partition of $treecomp(v)-\chi(v)$.
\end{lem}
\begin{proof} 
Let $C$ be a \component{v}\ such that 
$(C\cap W)\neq\emptyset$. We show that $C\subseteq W$.
Assume this is not true, i.e., $C-W \neq\emptyset$.
By definition of $treecomp(v)$, $\chi(T_v)=treecomp(v)\cup\chi(v)$.
Hence, any variable $Y\in (C-W)$ only occurs in the $\chi$ label
of vertices not belonging to $vertices(T_v)$.
However, $C$ is a \component{v}, therefore
$C\cap \chi(v)=\emptyset$. As a consequence,
$vertices(C,\HD)$ induces a disconnected subgraph of $T$, and thus 
contradicts Lemma~\ref{lem:connected}.  
\end{proof}

\begin{lem}\label{lem:treecomp}
Let $\HD=\tuple{T,\chi,\lambda}$ be an NF hypertree decomposition of 
a conjunctive query $Q$, and $r$ be a vertex of\/ $T$.
Then, $C=treecomp(s)$ for some child $s$ of $r$ if and only if $C$ is
an \component{r}\ of $Q$ and $C\subseteq treecomp(r)$.
\end{lem}
\begin{proof}
({\em If part.}) Assume $C$ is an \component{r}\ of $Q$ and
$C\subseteq treecomp(r)$. Let $children(r)$ denote the set of the vertices
of $T$ which are children of $r$.
Because $C\subseteq (treecomp(r)-\chi(r))$, $C$ must be included 
in $\bigcup_{s\in children(r)} \chi(T_s)$.
Moreover, for each subtree $T_s$ of $T$ such that $s\in children(r)$,
there is a (unique) \component{r}\ $treecomp(s)$ such that 
$\chi(T_s) = treecomp(s)\cup (\chi(s)\cap \chi(r))$.
Therefore, $C$ necessarily coincides with one of these components,
say $treecomp(\bar s)$ for some $\bar s \in children(r)$.

({\em Only if part.})
Assume $C = treecomp(s)$ for some child $s$ of $r$, and let
$C'=treecomp(r)$.
By definition of $treecomp(s)$, $C$ is an \component{r}, then
$(C\cap \chi(r))=\emptyset$.
Since $\HD$ is in normal form,
$\chi(T_s)= C\cup (\chi(s)\cap \chi(r))$ and $\chi(T_r)= (C'\cup \chi(r))$. 
Moreover, $s$ is a child of $r$, then 
$vertices(T_s)\subseteq vertices(T_r)$ and thus
$\chi(T_s)\subseteq \chi(T_r)$.
Therefore, $C\cup (\chi(s)\cap \chi(r)) \;\subseteq\; \chi(T_r)$,
and hence we immediately get $C\subseteq  (C'\cup \chi(r))$.
However, $(C\cap \chi(r))=\emptyset$ and thus $C\subseteq  C'$.
\end{proof}

\begin{lem}\label{lem:size}
For any NF hypertree decomposition $\HD=\tuple{T,\chi,\lambda}$ of a
query $Q$, $|vertices(T)|\leq |var(Q)|$ holds.
\end{lem}
\begin{proof} 
Follows from Lemma~\ref{lem:treecomp}, Lemma~\ref{lem:partition}, 
and Condition~2 of the normal
form, which states that, for any $v\in vertices(T)$, 
$\chi(v)\cap treecomp(v)\neq\emptyset$. Hence, 
$treecomp(v)-\chi(v)\subset treecomp(v)$ and thus,
for any child $s$ of $v$ in $T$, $treecomp(s)$
is actually a proper subset of $treecomp(v)$.
\end{proof}

\begin{lem}\label{lem:components}
Let $\HD=\tuple{T,\chi,\lambda}$ be an NF hypertree decomposition of a
query $Q$, $s$ a vertex of $T$, and $C$ a set of variables such that
$C\subseteq treecomp(s)$.
Then, $C$ is an \component{s}\ if and only if $C$ is a 
\component{var(\lambda(s))}.
\end{lem}
\begin{proof} 
Let $V= var(\lambda(s))$.
By Condition~4 of Definition~\ref{defn:hyper-decomp}, 
$(V\cap \chi(T_s))\subseteq \chi(s)$. 
Since $\HD$ is in normal form,
$V$ satisfies the following property.
\[ (1) \;\; (V\cap treecomp(s))\subseteq \chi(s). \]

({\em Only if part.})
Assume $C\subseteq treecomp(s)$ is an \component{s}.
From Property~1 above,  $C\cap V=\emptyset$ holds.
As a consequence, for any pair of variables $\{X,Y\}\subseteq C$, 
$X$ \adj{s}\ to $Y$ entails $X$ \adj{V}\ to $Y$.
Hence, $C$ is a \connected{V}\ set of variables.
Moreover, $\chi(s)\subseteq V$. Then, any \connected{V}\ set which is a
maximal \connected{s}\ set  is a maximal  \connected{V}\ set as well,
and thus $C$ is a \component{V}.

({\em If part.})
Assume $C\subseteq treecomp(s)$ is a \component{V}. 
Since $\chi(s)\subseteq V$, $C$ is clearly \connected{s}.
Thus, $C\subseteq C'$, where $C'$ is an \component{s}\ and,
by Lemma~\ref{lem:partition}, $C'\subseteq (treecomp(s)-\chi(s))$ holds.
By the ``only if'' part of this lemma, $C'$ is a \component{V},
therefore $C$ cannot be a proper subset of $C'$, and  $C'=C$ actually holds.
Thus, $C$ is an \component{s}.
\end{proof}

\subsection{A LOGCFL Algorithm Deciding $k$-bounded Hypertree-Width}
\label{sec:main}

Figure~\ref{fig:algo-kb} shows 
the algorithm {\tt k-decomp},
deciding whether a given conjunctive
query $Q$ has a $k$-bounded hypertree-width decomposition.
In that figure, 
we give  a  high level  description of an alternating algorithm, to
be run on an alternating Turing machine (ATM).
The details of how the algorithm can be effectively implemented on a
logspace ATM will be given later (see Lemma~\ref{lem:atm}).

To each computation tree $\tau$ of {\tt k-decomp} on input query $Q$, 
we associate a hypertree  $\delta(\tau)=\tuple{T,\chi,\lambda}$, called
the {\em witness tree} of $\tau$, defined as follows:
For any existential configuration of $\tau$ corresponding to the
``guess'' of some set $S\subseteq atoms(Q)$
during the computation of {\em k-decomposable($C,R$)},
for some \component{var(R)}\ $C$,
(i.e., to Step 1 of \kdecomp), $T$ contains a vertex $s$.
In particular, the vertex $s_0$ guessed at the initial call
{\em k-decomposable($var(Q),\emptyset$)}, is the root of $T$.

There is an edge between vertices $r$ and $s$ of $T$, where $s\neq s_0$,
if $S$ is guessed at Step 1 during the computation of 
{\em k-decomposable($C,R$)}, for some \component{var(R)}\ $C$
($S$ and $R$ are the (guessed) sets of atoms of $\tau$ corresponding
to $s$ and $r$ in $T$, respectively).
We will denote $C$ by $comp(s)$, and $r$ by $father(s)$.
Moreover, for the root $s_0$ of $T$, we define $comp(s_0)=var(Q)$.

The vertices of $T$ are labeled as follows.
$\lambda(s)=S$
(i.e., $\lambda(s)$ is the guessed set $S$ of atoms corresponding to $s$),
for any vertex $s$ of $T$. If $s_0=root(T)$, let
$\chi(s_0)= var(\lambda(s_0))$; for any other vertex $s$, let
$\chi(s)= var(\lambda(s))\ \cap\ ( \chi(r) \cup C)$,
where $r = father(s)$ and $C=comp(s)$.

\begin{figure}
\begin{center}
\fbox{\begin{minipage}{17.5cm}
{ 
\begin{tabbing}
{\bf ALTERNATING ALGORITHM} \kdecomp\ \\
{\bf Input:} A non-empty Query $Q$.\\
{\bf Result:} ``Accept'', if $Q$ has $k$-bounded hypertree width;
``Reject'', otherwise. \\
\ \\
{\bf Procedure} {\em k-decomposable($C_R$: SetOfVariables, $R$: SetOfAtoms)}\\
{\bf beg}\={\bf in}\\
1)\> {\bf Guess} a set $S\subseteq atoms(Q)$ 
                                     of $k$ elements at most;\\
2)\> {\bf Check} that all the following conditions hold:\\
\> 2.a) \ \ \=  $\forall P\in atoms(C_R),\; 
 (var(P)\cap var(R))\subseteq var(S)$ and\\ 
\> 2.b) \> $var(S) \cap C_R \; \neq \emptyset$\\
3) \> {\bf If} the check above fails {\bf Then} {\bf Halt} and {\bf
Reject}; {\bf Else}\\
  \> Let ${\cal C} :=\{ C\subseteq var(Q)\;|\;C\ is\
a\;$ {\em \component{var(S)} and} $C\subseteq C_R  \}$;\\
4)\> {\bf If, for each} $C\in {\cal C}$,\ \ {\em k-decomposable}$(C,S)$\\
  \> \ \ \ {\bf Then Accept}\\
  \> \ \ \ {\bf Else Reject}\\
{\bf end;}\\
\ \\
{\bf begin}{\em (* MAIN *)}\\
\> {\bf Accept} if {\em k-decomposable}$(var(Q),\emptyset)$\\
{\bf end.}\\
\end{tabbing}
}
\end{minipage}}
\end{center}
\vspace{-0.4cm} 
\caption{\em A non-deterministic algorithm 
             deciding $k$-bounded hypertree-width}
\label{fig:algo-kb}
\vspace{-0.4cm}
\end{figure}

\begin{lem}\label{lem:algo-complete}
For any given query $Q$ such that $hw(Q)\leq k$,
 \kdecomp\  accepts $Q$.
Moreover, for any $c\leq k$,
each $c$-width hypertree-decomposition of $Q$ in normal form
is equal to some witness tree for $Q$.
\end{lem}
\begin{proof}
Let $\HD=\tuple{T,\chi,\lambda}$ be a $c$-width NF hypertree decomposition of 
a conjunctive query $Q$, where $c\leq k$. 
We show that there exists an accepting computation tree $\tau$ for
\kdecomp\ on input query $Q$ such that
$\delta(\tau)=\tuple{T',\chi',\lambda'}$  ``coincides'' with
$\HD$. Formally, there exists a bijection $f: vertices(T)\rightarrow
vertices(T')$ such that, for any pair of vertices $p,q\in T$,
$p$ is a child of $q$ in $T$ iff $f(p)$ is a child of $f(q)$ in
$T'$, $\lambda(p)=\lambda'(f(p))$, $\lambda(q)=\lambda'(f(q))$,
$\chi(p)=\chi'(f(p))$, and $\chi(q)=\chi'(f(q))$.

To this aim, we impose to \kdecomp\ on input $Q$ the following choices
of sets $S$ in Step~1:
\begin{itemize}
\item[a)] For the initial call {\em $k$-decomposable}$(var(Q),\emptyset)$, 
the set $S$ chosen in Step~1 is $\lambda(root(T))$.
\item[b)] Otherwise, for a call {\em $k$-decomposable}$(C_R,R)$,
if $R$ is the label $\lambda(r)$ of some vertex $r$, and if $r$ has a
child $s$ such that $treecomp(s)=C_R$, then choose $S=\lambda(s)$ in
Step~1.
\end{itemize}
We use structural induction on trees to prove that, 
for any vertex $r\in vertex(T)$, if we denote $f(r)$ by $r'$, the following 
equivalences hold:
$\lambda(r)=\lambda'(r')$; $treecomp(r)=comp(r')$; and $\chi(r)=\chi'(r')$.

{\em Basis:\/}  For $r=root(T)$, we set
$f(root(T)) := root(T')$. Thus,
by choosing $\lambda'(f(r))=\lambda(r)$ 
as described at Point a) above, all the equivalences trivially hold.

{\em Induction Step:} 
Assume that the equivalence holds for some vertex $r\in vertices(T)$.
Then, we will show that the statement also holds for every child of $r$.
Let $r'\in vertices(T')$ denote $f(r)$, and
let $s$ be any child of $r$ in $T$.
By Lemma~\ref{lem:treecomp} and Lemma~\ref{lem:components},
the \component{r}\ $treecomp(s)$ coincides
with some \component{var(\lambda'(r'))}\ $comp(s')$ corresponding to the call 
{\em $k$-decomposable}$(comp(s'),\lambda'(r'))$ that generated 
a child $s'$ of $r'$, which we define to be the image of $s$,
i.e., we set $f(s) := s'$.
Since $\HD$ is a $k$-width hypertree decomposition, 
and the induction hypothesis holds, it easily follows
that, by choosing $\lambda(s)=\lambda'(s')$ as prescribed at Point b) above, 
no check performed in Step~2 of the call 
{\em $k$-decomposable}$(comp(s'),\lambda'(r'))$ can fail.

Next we show that $\chi(s)=\chi'(s')$.
Let $C= comp(s')=treecomp(s)$, and $V=var(\lambda(s))=var(\lambda'(s'))$.
By Condition~4 of Definition~\ref{defn:hyper-decomp}, 
$V\cap \chi(T_s)\;\subseteq\; \chi(s)$ holds.
Since $\HD$ is in normal form, we can replace $\chi(T_s)$ by 
$C\cup \chi(s)$ according to Condition~1 of Definition~\ref{defn:nf}, 
and we get $V\cap (C\cup \chi(s)) \;\subseteq\; \chi(s)$.  
Hence, we obtain the following property 
$$(1)\;\; V\cap C \;\subseteq\; \chi(s).$$

Now, consider $\chi'(s')$. By definition of witness tree,
$\chi'(s')=\; V \cap (\chi'(r') \cup C) \;=\; V\cap (\chi(r)\cup C)$. 
By Property $(1)$ above, $V\cap C \;\subseteq\; \chi(s)$.
Moreover, $\HD$ is in NF, and Condition~3 of the normal form entails
that $(V\cap \chi(r))\subseteq \chi(s)$. As a consequence,
$\chi'(s') \subseteq \chi(s)$.
We claim that this inclusionship  cannot be proper.
Indeed, by definition of $\chi'(s')$, 
if $\chi'(s') \subset \chi(s)$, there exists a 
variable $Y\in\chi(s)$ which belongs neither to $\chi(r)$, nor to $C$.
However, this entails that $Y$ belongs to some other \component{r}\
and thus $s$ violates Condition~1 of the normal form.  

In summary, \kdecomp\ accepts $Q$ 
with the accepting computation tree $\tau$ determined by the choices
described above, and its witness tree $\delta(\tau)$ is a $c$-width hypertree
decomposition of $Q$ in normal form.
\end{proof}

\begin{lem}\label{lem:comp-algo}
Assume that \kdecomp\ accepts an input query $Q$ with an accepting
computation tree $\tau$ and let $\delta(\tau)=\tuple{T,\chi,\lambda}$
be the corresponding witness tree. Then, for any vertex $s$ of $T$: 
\begin{itemize}
\item[a)]
if $s\neq root(T)$, then $comp(s)$ is a \component{father(s)};
\item[b)]
for any $C\subseteq comp(s)$, 
$C$ is an \component{s}\ if and only if $C$ is a \component{var(\lambda(s))}.
\end{itemize}
\end{lem}
\begin{proof} 
We use structural induction on the tree $T$.

{\em Basis:\/} Both parts of the lemma trivially holds 
if $s$ is the root of $T$. In fact, in this case, 
we have $\chi(s)=var(\lambda(s))$, by definition of witness tree.

{\em Induction Step:} 
Assume that the lemma holds for some vertex $r\in vertices(T)$.
Then, we will show that both parts hold for every child of $r$.
The induction hypothesis states that 
any \component{var(\lambda(r))}\ included in $comp(r)$ is an
\component{r}\ included in $comp(r)$, and vice versa.
Moreover, if $r\neq root(T)$, then $comp(r)$ is a 
\component{father(r)}; otherwise, i.e., $r$ is the root, 
$comp(r)=var(Q)$, by definition of witness tree.
Let $s\in vertices(T)$ be a child of $r$ and let $V= var(\lambda(s))$. 
We first observe that, by definition of the variable labeling $\chi$ of
the witness tree, it follows that 
$var(\lambda(s))\cap comp(s) \subseteq \chi(s)$. 
Hence, the following holds. \\
\underline{Fact 1}: $(V-\chi(s))\cap comp(s) =\emptyset$. \\
({\em Point a.})
Immediately follows by the definition of $comp(s)$ and by the 
induction hypothesis. Indeed, $r$ is the father of $s$ and by
the induction hypothesis any 
\component{var(\lambda(r))}\ included in $comp(r)$ is an
\component{r}\ included in $comp(r)$. Thus, in particular,
$comp(s)$ is an \component{r}. \\
({\em Only if part, Point b.})
Assume that a set of variables 
$C\subseteq comp(s)$ is an \component{s}.
By Fact 1, $C\cap (V-\chi(s))=\emptyset$ holds, and
for any pair of variables $\{X,Y\}\subseteq C$, 
$X$ \adj{s}\ to $Y$ entails $X$ \adj{V}\ to $Y$.
Hence, $C$ is a \connected{V}\ set of variables.
Moreover, $\chi(s)\subseteq V$. Then, any \connected{V}\ set which is 
a maximal \connected{s}\ set  is a maximal  \connected{V}\ set as well,
and thus $C$ is a \component{V}.\\
({\em If part, Point b.})
We proceed by contradiction.
Assume $C\subseteq comp(s)$ is a \component{V}, but 
$C$ is not an \component{s},
i.e., $C$ is not a maximal \connected{s}\ set of variables.
Since $\chi(s)\subseteq V$, $C$ is clearly \connected{s}, then
it is not maximal. That is,
there exists a pair of variables $X\in C$ and $Y\not\in C$ such
that $X$ is \adj{s}\ to $Y$, but $X$ is not \adj{var(\lambda(s))}\ to $Y$.
Let $A$ be any atom proving their adjacency w.r.t. $s$, i.e.,
$\{X,Y\}\subseteq var(A)-\chi(s)$. Hence, because $X\in C$ and
$X$ is not \adj{V}\ to $Y$, it follows that $Y\in (V-\chi(s))$.
By Fact 1,
$(V-\chi(s))\cap comp(s)=\emptyset$, therefore $Y\not\in comp(s)$.
In summary, $X\in comp(s)$ and $Y\not\in comp(s)$.
Moreover, $comp(s)\subseteq comp(r)$, by Step~4 of \kdecomp.
Hence, by induction hypothesis, $comp(s)$ is an \component{r}\ and
thus $X$ is not \adj{r}\ to $Y$. Consider again the atom $A$.
We get $\{X,Y\}\not\subseteq var(A)-\chi(r)$. Since $X\in comp(s)$,
the variable $Y$ must belong to $\chi(r)$.
However, by definition of witness tree, $Y\in \chi(r)$ and
$Y\in var(\lambda(s))$ entail that $Y\in \chi(s)$, which is a contradiction.
\end{proof}

\begin{lem}\label{lem:z}
Assume that \kdecomp\ accepts an input query $Q$ with an accepting
computation tree $\tau$. Let $\delta(\tau)=\tuple{T,\chi,\lambda}$
be the corresponding witness tree, and 
$s\in vertices(T)$.
Then, for each vertex $v\in T_s :$
\[
\begin{array}{lll}
\chi(v) & \subseteq & comp(s)\cup \chi(s)\\
comp(v) & \subseteq & comp(s)
\end{array}
\]
\end{lem}
\begin{proof}
We use induction on the distance $d(v,s)$ between any vertex 
$v\in vertices(T_s)$ and $s$.
The basis is trivial, since   $d(v,s)=0$ means $v=s$.

{\em Induction Step.} Assume both statements hold for distance
$n$. Let $v\in vertices(T_s)$ be a vertex such that $dist(v,s)=n+1$.
Let $v'$ be the father of $v$ in $T_s$. Clearly, $dist(v',s)=n$,
thus 
\begin{itemize}
\item[($a$)] $\chi(v')\subseteq (comp(s)\cup \chi(s))$; and
\item[($b$)] $comp(v')\subseteq comp(s)$.
\end{itemize}
$v$ is generated by some call {\em $k$-decomposable}$(comp(v),\lambda(v'))$.
By the choice of $v$ and the definition of witness tree, it must hold 
($a'$) $\chi(v)\subseteq (comp(v)\cup \chi(v'))$, and
by Step~4 of the call {\em $k$-decomposable}$(comp(v'),\lambda(father(v')))$
we get ($b'$) $comp(v)\subseteq comp(v')$.
By ($a'$) and ($b'$),
we obtain ($a''$) $\chi(v)\subseteq (comp(v')\cup \chi(v'))$.
By ($a''$), ($b$), and ($a$) we get $\chi(v)\subseteq (comp(s)\cup \chi(s))$.
Moreover, ($b$) and ($b'$) yield $comp(v)\subseteq comp(s)$.  
\end{proof}

\begin{lem}\label{lem:algo-sound}
If  \kdecomp\  accepts an input query $Q$,
then $hw(Q)\leq k$.
Moreover, each witness tree for $Q$
is a  $c$-width hypertree-decomposition of $Q$ in normal form, 
where $c\leq k$.
\end{lem}
\begin{proof}  
Assume that $\tau$ is an accepting computation tree of \kdecomp\ on
input query $Q$. We show that $\delta(\tau)=\tuple{T,\chi,\lambda}$ is an
NF $c$-width hypertree decomposition of $Q$, for some $c\leq k$.

First, we will prove that $\delta(\tau)$ fulfils all the properties
of Definition~\ref{defn:hyper-decomp} and is thus a
hypertree decomposition of $Q$.

$\underline{\mbox{\em Property 1}:}$\ 
$\forall A\in atoms(Q) \; \exists v\in vertices(T) \;s.t.\;
var(A)\subseteq \chi(v)$.

We first prove the following claim.

CLAIM A: 
Let $s$ be any vertex of $T$, and 
let $C_r=comp(s)$.
Then, for each $P\in atoms(C_r)$ it holds that
\begin{itemize}
\item[(a)]
$\forall A\in (atoms(Q)-atoms(C_r)), \; 
            (var(P)\cap var(A))\subseteq \chi(s)$; and 
\item[(b)]
either $var(P)\subseteq \chi(s)$ or 
there exists an \component{s}\ 
$C_s\subseteq C_r$ such that $P\in atoms(C_s)$.
\end{itemize}
\begin{quote}
{\bf Proof of Claim A.}
{\em (Part a).}
We use structural induction on the tree $T$.\\
{\em Basis:\/} Part (a) of the claim trivially holds 
if $s$ is the root of $T$. In fact, in this case, we have $C_r=comp(s)=var(Q)$
and hence $atoms(C_r)=atoms(Q)$.\\
{\em Induction Step:} 
Assume that Part (a) holds for some vertex $s\in vertices(T)$.
Then, we will show that the statement also holds for every child of $s$.
Let $C_r=comp(s)$ and $V=var(\lambda(s))$. 
The induction hypothesis states that
$\forall P\in atoms(C_r)$ and
$\forall A\not\in atoms(C_r), \; var(P)\cap var(A)\subseteq \chi(s)$.
By Step~4 of \kdecomp, for each \component{V}\ $C$ s.t. $C\subseteq C_r$, 
$T$ contains a vertex $q$ such that $comp(q)=C$ and $father(q)=s$.
Moreover, by Lemma~\ref{lem:comp-algo}, $C$ is an \component{s}.
Let $C_s$ be an \component{s}\ s.t. $C_s\subseteq C_r$, and
let $P'$ belong to $atoms(C_s)$. 
By choice of $C_s\subseteq C_r$, $P'$ also belongs to $atoms(C_r)$.
First note that, $\forall A \not\in atoms(C_s)$, we have 
\[(1)\;\; var(P')\cap var(A) \subseteq \chi (s).\] 
Indeed, if $var(A)\subseteq \chi(s)$ (1) is trivial, and 
if $A\not\in atoms(C_r)$, it
follows from the induction hypothesis. Otherwise, i.e. if $A$ 
contains some variable belonging to
another \component{s}\ included in $C_r$, it immediately follows by
definition of \component{s}.
Now, for the component $C_s$, Step~1 of 
{\em $k$-decomposable}$(C_s,\lambda(s))$ guesses a vertex $s'$ 
such that, $\forall B\in atoms(C_s)$,  
$(var(B)\cap var(\lambda(s))) \subseteq var(\lambda(s'))$.
In particular, $(var(P')\cap var(\lambda(s))) \subseteq var(\lambda(s'))$.
Because $\chi(s)\subseteq var(\lambda(s))$,
this yields $(var(P')\cap \chi(s)) \subseteq (var(\lambda(s'))\cap\chi(s))$.
By definition of witness tree, 
$(var(\lambda(s'))\cap\chi(s))\subseteq\chi(s')$, hence we get
$(var(P')\cap \chi(s)) \subseteq \chi(s')$.
By combining this result with relationship~(1) above, we get that
$\forall A \not\in atoms(C_s)$ 
$(var(P')\cap var(A)) \subseteq (var(P')\cap \chi(s))\subseteq \chi(s')$.
Hence, Part (a) of the claim holds even for $s'$ and thus for
every child of $s$ in $T$.

{\em (Part b).} 
Let $s$ be any vertex of $T$, 
let $C_r=comp(s)$, and let $P$ belong to $atoms(C_r)$. 
Assume that $var(P)\not\subseteq \chi(s)$ and that 
$P\in atoms(C'_s)$, where $C'_s$ is an
\component{s}\ 
not included in $C_r$, i.e., $C'_s\not\subseteq C_r$.
Then, there exists a variable $Y\in C'_s$ s.t. $Y\not\in C_r$ and there
is an \spath{s}\  $\pi$ from $Y$ to any variable $X\in (var(P)-\chi(s))$. 

Let $A$ be an atom belonging to both $atoms(C_r)$ and $atoms(C'_s)$.
Then, $var(A)-\chi(r)\subseteq C_r$ and $var(A)-\chi(s)\subseteq C_s$ hold.
As a consequence, $(var(A)\cap C'_s)\subseteq C_r$.
Indeed, if this is not true, there exists a variable $Z\in(var(A)-\chi(s))$
such that $Z\in\chi(r)$.
By the definition of the $\chi$ labeling of a witness tree, this entails
that $Z\in (var(A)\cap var(\lambda(r)) )$, but $Z\not\in var(\lambda(s))$,
which contradicts the fact that 
$A$ satisfies the condition checked at Step 2.a of \kdecomp,
because $\tau$ is an accepting computation tree.

Therefore, there exist two atoms $\{Q',P'\}\subseteq atoms(\pi)$
belonging to $atoms(C'_s)$ and adjacent in $\pi$ s.t.
$Q'\not\in atoms(C_r)$, $P'\in atoms(C_r)$, 
and $var(Q')\cap var(P') \not\subseteq \chi(s)$. 
However, this contradicts  Part (a) of the claim. $\diamond$
\end{quote}

Note that, by Lemma~\ref{lem:comp-algo}, in the Step~4 of any
call {\em $k$-decomposable}$(C,\lambda(r))$ of an accepting
computation of \kdecomp, \component{s}s included in $C$ and 
\component{var(\lambda(s))}s included in $C$ coincide.
Thus, Property~1 follows by inductive application of Part b of Claim A.
In fact, Part (b) of the claim applied to the root $s_0$ of $T$,
states that, $\forall A\in atoms(Q)$, either $var(A)\subseteq \chi(s_0)$, 
or $A\in atoms(C_S)$ for some \component{S}\ $C_S$ of $Q$ that will be
further treated in Step~4 of the algorithm.
Thus, $var(A)$ is covered eventually by some chosen set of atoms $S$,
i.e., there exists some vertex $s$ of $T$, such that
$\lambda(s)=S$, and $var(A)\subseteq \chi(s)$.

$\underline{\mbox{\em Property 2}:}$\ 
{\em For each variable $Y\in var(Q)$ the set $\{ v\in vertices(T)\;|\: Y\in
\chi(v)\}$ induces a connected subtree of $T$. }

Assume that Property~2 does not hold. Then, there exists a variable
$Y\in var(Q)$ and two vertices $v_1$ and $v_2$ of $T$ such that 
$Y\in (\chi(v_1)\cap \chi(v_2))$ but the unique path from $v_1$ to $v_2$
in $T$ contains a vertex $w$ such that $Y\not\in \chi(w)$.
W.l.o.g, assume that $v_1$ is adjacent to $w$ and that $v_2$ is a descendant
of $w$ in $T$, i.e., $v_2\in vertices(T_w)$.
There are two possibilities to consider:
\begin{itemize}
\item \underline{$v_1$ is a child of $w$} and $v_2$ belongs
to the subtree $T_p$ of another child $p$ of $w$.
However, this would mean that, by Step~4 of \kdecomp\ and by
Lemma~\ref{lem:z}, the variables in sets $V_1= (\chi(v_1)-\chi(w))$ and  
$V_2= (\chi(v_2)-\chi(w))$ belong to distinct \component{w}s.
But this is not possible, because $Y\in (V_1\cap V_2)$.
\item \underline{$w$ is a child of $v_1$} and $v_2$ belongs
to the subtree $T_w$ of $T$ rooted at $w$.
Then, $\lambda(w)$ was chosen as set $S$ in Step~1 of
$k$-decomposable$(C,\lambda(v_1))$, where $C$ is a \component{v_1}.
Note that $Y\in \chi(v_1)$ entails $Y\not\in C$, by definition of
 \component{v_1}.
Since $v_2$ belongs to the subtree $T_w$, by Lemma~\ref{lem:z} it
holds that $\chi(v_2)\subseteq (C\cup \chi(w))$.
This is a contradiction, because $Y\in \chi(v_2)$, but $Y$ belongs
neither to $\chi(w)$, nor to $C$.
\end{itemize}

$\underline{\mbox{\em Property 3}:}$\ 
$\forall p\in vertices(T), \; \chi(p)\subseteq var(\lambda(p))$.

Follows by definition of the $\chi$ labeling of a witness tree.

$\underline{\mbox{\em Property 4}:}$\ 
$\forall p\in vertices(T), \; var(\lambda(p))\cap \chi(T_p)\subseteq \chi(p)$.

Let $v$ be any vertex in $T_p$, and let $V=var(\lambda(p))$.
By Lemma~\ref{lem:z}, $\chi(v)\subseteq comp(p)\cup \chi(p)$.
Hence, $V\cap \chi(v) \subseteq \; V\cap comp(p) \cup \chi(p)$,
because Property~3 holds for $p$.
However, by definition of witness tree, $(V\cap comp(p))\subseteq \chi(p)$,
and thus $(V\cap \chi(v))\subseteq \chi(p)$.

Thus, $\delta(\tau)$ is a hypertree decomposition of $Q$. 
Let $c$ be the width of $\delta(\tau)$.
Since Step~1 of \kdecomp\ only chooses set of atoms having cardinality
bounded by $k$, $c\leq k$ holds. 

Moreover,  $\delta(\tau)$ is in normal form. Indeed,
Condition~2 and Condition~3 of Definition~\ref{defn:nf} 
hold by Step~2.b of \kdecomp, and by definition of the
$\chi$ labeling of a witness tree.
Finally, since $\delta(\tau)$ is a hypertree decomposition,
by Lemma~\ref{lem:path}, Lemma~\ref{lem:z}, and the definition of the
$\chi$ labeling of a witness tree,
we get that Condition~1 holds for $\delta(\tau)$, too. 
\end{proof}

By combining Lemma~\ref{lem:algo-complete} and
Lemma~\ref{lem:algo-sound} we get:

\begin{theo}\label{theo:algo-correct}
\kdecomp\ accepts an input query $Q$ if and only if $hw(Q)\leq k$.
\end{theo}

\begin{lem}\label{lem:atm}
\kdecomp\  can be implemented on a logspace ATM having
polynomially bounded tree-size.
\end{lem}
\begin{proof}  
Let us refer to logspace ATMs with polynomially bounded tree-size 
as $\LCFL$-ATMs. We will outline how the algorithm 
\kdecomp\ can be implemented on an $\LCFL$-ATM $M$.

We first describe the data-structures used by $M$. 
Instead of manipulating atoms directly,
{\em indices} of atoms will be used in order
to meet the logarithmic space bound.
Thus the $i$-th atom occurring in the given
representation of the input query $Q$ will be 
represented by integer $i$. 
Sets 
of at most $k$ atoms, $k$-sets for short, are represented by $k$-tuples of 
integers; since $k$ is fixed, representing such sets
requires logarithmic space only. 
Variables are 
represented as integers, too.

If $R$ is a $k$-set, then a \component{var(R)}\ $C$ is represented by
a pair $\langle rep(R), first(C)$, where $rep(R)$ is 
the representation of the $k$-set $R$, and 
$first(C)$ is the smallest integer representing a variable 
of the component $C$. For example, the $\emptyset$-component 
$var(Q)$ is represented by the pair $\langle
rep(\emptyset),1\rangle$. It is thus clear that \component{var(R)}s
can be represented in logarithmic space, too.

The main data structures carried with each configuration of $M$ 
consist of (the representations of): 
\begin{itemize}
\item a $k$-set $R$,\footnote{The separate representation of $R$ is 
actually slightly redundant, given that $R$ also occurs in the
description
of the $[R]$-component $C_R$.}
\item a \component{var(R)}\ $C_R$,
\item a $k$-set $S$, and 
\item a \component{var(S)}\  $C$.
\end{itemize}
Not all these items will contain useful data in all configurations.
We do not describe further auxiliary logspace data structures that may be 
used for control tasks and for other tasks such as counting  
or for performing some of the SL subtasks described below.

We are now ready to give a  description of 
the computation $M$ performs on an input query $Q$.

To facilitate the description,
we will specify some subtasks of the computation, that are 
themselves solvable in LOGCFL, as macro-steps without 
describing their corresponding computation (sub-)trees.
We may imagine a macro-step as a special kind of configuration 
-- termed {\em oracle configuration} -- 
that acts as an oracle for the subtask to be solved. 

Each oracle configuration can be {\em normal} or {\em converse}.

A {\em normal} oracle configuration has the following effect.
If the subtask is negative, this configuration has no children 
and amounts to a 
REJECT. Otherwise, its value (ACCEPT or REJECT) 
is identical to the value 
of its unique successor configuration. 

A {\em converse} oracle configuration has the following effect.
If the subtask is negative, this configuration has no children 
and amounts to an
ACCEPT. Otherwise, its value (ACCEPT or REJECT) 
is identical to the value 
of its unique successor configuration. 

From the definition of logspace ATM with polynomial tree-size,
it follows that 
any polynomially tree-sized logspace ATM $M$ with $\LCFL$ oracle 
configurations (where an oracle configuration contributes
1 to the size of an accepting subtree)  
is equivalent to a standard logspace ATM having 
polynomial tree size. 

$M$ is started with $R$  initialized
to the empty set and $C_R$ having value  $var(Q)$. 

We describe the evolution of $M$ corresponding to a procedure 
call {\em k-decomposable$(C_R,R)$}.

Instruction 1 is performed by  guessing an arbitrary 
$k$-set $S$ of atoms.

The ``Guess'' phase of Instruction 1 is implemented by an
existential configuration of the ATM. (Actually, 
it is implemented by a subtree of 
existential configurations, given that a single existential
configuration
can only guess one bit; note however, that 
each accepting computation tree will contain 
only one branch of this 
subtree.)

Checking Step~2 is in symmetric logspace ($\SL$).
The most difficult task is to enumerate atoms of 
$atoms(C_R)$, which in turn -- as most substantial 
subtask -- requires to enumerate
the variables of $C_R$.
Remember that $C_R$ is given in the form $\langle rep(R), i\rangle$
as described above. Thus, enumerating $C_R$ amounts to 
cycling over all variables $j$ and checking whether $j$ is 
$[R]$-connected to $i$. The latter subtask is easily seen to 
be in SL because it essentially amounts to a connectedness-test 
of two vertices in an
undirected graph. It follows that the entire checking-task 
of Instruction 2 is in $\SL$. Since $\SL\subseteq\LCFL$, this
correspond to a $\LCFL$-subtask. We can thus assume that the
checking-task is  performed by some normal oracle configuration.
If the oracle computation fails at some branch, the branch ends in a REJECT,
otherwise, the guessed $k$-set $S$ corresponding to 
that branch satisfies all the conditions checked by Step~2 of \kdecomp.

Steps 3 and 4 together intuitively correspond to a 
``big'' universal configuration that universally quantifies 
over all subtrees corresponding to the procedure calls 
{\em k-decomposable$(C,S)$} for all $C\in {\cal C}$. 
This could be realized as follows.
First, a subtree of universal configurations 
enumerates all candidates $C_i=\langle rep(S),i\rangle$ 
for $1\leq i\leq |var(Q)|$, for \component{var(S)}s. Each branch 
of this subtree (of polynomial depth) computes exactly one 
candidate $C_i$. Each such branch is 
expanded by a converse oracle  configuration 
checking whether $C_i$ is effectively a \component{var(S)}\ contained 
in $C_R$. Thus, branches that do not correspond to such a 
component are terminated with an ACCEPT configuration
(they are of no interest), while 
all other branches are further expanded. Each branch $C_i$ 
of the latter type is expanded by the subtree corresponding to 
the recursive call {\em k-decomposable$(C_i,S)$}. 

We have thus completely described a logspace ATM $M$ with oracle 
configurations that implements \kdecomp. 
It is easy to see that this machine has polynomial accepting
computation trees. In fact, this is seen from the fact that 
there exist only a polynomial number of choices for set $S$ in Step 1,
and that no such set is chosen twice in any accepting computation
tree. 
\end{proof}

From the lemmas above and Proposition \ref{prop:atm}, we obtain the following theorem.

\begin{theo}\label{theo:logcfl}
Deciding whether a conjunctive query $Q$ has $k$-bounded hypertree-width
is in $\LCFL$.
\end{theo}

In fact, the following proposition states that an accepting computation tree
of a bounded-treesize logspace ATM can be {\em computed} in
(the functional version of) $\LCFL$.
\begin{prop}[\cite{gott-etal-98c}]
\label{prop:certificates}
Let $M$ be a bounded-treesize logspace ATM 
recognizing a language $A$.
It is possible to construct a $\LLOG$ transducer $T$ 
which for each input $w\in A$ outputs 
a single (polynomially-sized) 
accepting tree for $M$ and $w$.
\end{prop}

By Lemma \ref{lem:algo-complete} and Lemma \ref{lem:algo-sound},
we have a one-to-one correspondence between the 
NF k-width hypertree decompositions and accepting computation
trees of \kdecomp.
Thus, by Proposition~\ref{prop:certificates}, we get that hypertree
decompositions are efficiently {\em computable}.

\begin{theo}\label{theo:computation}
Computing a $k$-bounded hypertree decomposition (if any)
of a conjunctive query $Q$ is in $\FCFL$, i.e., in functional $\LCFL$.
\end{theo}

Since $\LCFL$ is closed under $\FCFL$ reductions \cite{gott-etal-98c},
the two following statements follow from the theorem above
and Theorem \ref{th:hdcfl} and Theorem \ref{th:hdpoly},
respectively.

\begin{coroll}
Deciding whether a k-bounded hypertree-width query $Q$ evaluates
to {\em true} on a database $\DB$ is $\LCFL$-complete.
\end{coroll}

\begin{coroll}
The answer of a (non-Boolean) k-bounded hypertree-width query $Q$
can be {\em computed} in time polynomial in the combined size of
the input instance and of the output relation.
\end{coroll}

\subsection{A Datalog Program Recognizing Queries of
$k$-bounded Hypertree-Width}

In this section we show a straightforward polynomial-time
implementation of the LOGCFL algorithm above.
In particular, we reduce (in polynomial time) the problem of deciding
whether there exists a $k$-bounded hypertree-width decomposition of a
given conjunctive query $Q$ to the problem of solving a Datalog
program.

First, we associate an identifier (e.g., some constant number)
to each $k$-vertex (non empty subset of $Q$ consisting of $k$ atoms at
most) $R$ for $Q$, and
to each \component{R}\ $C$ for any $k$-vertex $R$ of $Q$.
Moreover, we have a new identifier $root$ which intuitively will be
the root of any possible tree-decomposition, and a new identifier $varQ$
which encodes the set of 
all the variables of the query and hence is seen as a component
including any subset of $var(Q)$.

Then, we compute the following relations:\footnote{For the sake of
clarity, we directly refer to objects by means of their associated
identifiers (which we also use as logical terms in the program).}
\begin{itemize}
\item {\em k-vertex($\cdot$)}: Contains a tuple $\tuple{R}$ 
  for each  {\em k-vertex} $R$ of $Q$.
\item {\em component($\cdot,\cdot$)}: Contains a tuple $\tuple{C_R,R}$ for each
  \component{R}\ $C_R$ of some  $k$-vertex $R$. \\
  Moreover, it contains the  tuple $\tuple{varQ,root}$.
\item {\em meets-condition($\cdot,\cdot,\cdot$)}: Contains any tuple
  $\tuple{S,R,C_R}$ s.t. $S$ and $R$ are $k$-vertices, $C_R$ is an
  \component{R}, and the following conditions hold:
  $var(S) \cap C_R\neq\emptyset$, 
  and $\forall P\in atoms(C_R)$ $var(P)\cap  var(R)\subseteq var(S)$. \\
  Moreover, it contains a tuple $\tuple{S,root,varQ}$ for any $k$-vertex $S$.
\item {\em subset($\cdot,\cdot$)}: Encodes the standard set-inclusion
  relationship between \component{R}s of $Q$.
\end{itemize}

Let $\cal P$ be the following Datalog program:
\begin{enumerate}\em
\item $k$-decomposable($R,C_R$)$\derives$
           $k$-vertex(S),  meets-conditions($S,R,C_R$),
$\neg$ undecomposable($S,C_R$)
\item undecomposable($S,C_R$)$\derives$ component($C_S,S$),
                       subset($C_S,C_R$), $\neg$ $k$-decomposable($S,C_S$).
\end{enumerate}

It is easy to see that $hw(Q)\leq k$ 
if and only if ${\cal P}\models$ {\em $k$-decomposable($root,varQ$)}.

Note that ${\cal P}$ is locally stratified on the base relations 
to which it is applied, and it is clearly evaluable in polynomial time.

\section{Bounded Hypertree-Width vs Related Notions}

Many relevant cyclic queries are -- in a precise sense --
close to acyclic queries because they can be decomposed 
via  low bandwidth decompositions to  acyclic queries.

The main classes of bounded-width queries  
considered in database theory and in artificial intelligence  
are the following:

\begin{itemize}
\item {\em Queries of bounded treewidth.}
Treewidth is the best-known graph theoretic measure of tree-similarity.
The concept of treewidth is based on the notion of {\em tree-decomposition}
of a graph. The concept of treewidth is easily
generalized to hypergraphs and thus to conjunctive queries. 
Conjunctive queries of bounded treewidth can be answered in 
polynomial time~\cite{chek-raja-97}. For each fixed $k$, 
deciding whether a query has treewidth $k$ is in $\LCFL$.

\item {\em Queries of bounded degree of cyclicity}.
This concept was introduced by Gyssens
et al.~\cite{gyss-para-84,gyss-etal-94}
and is based on the notion of {\em hinge-tree decomposition}. 
The smaller the degree of cyclicity of a hypergraph, the more the hypergraph 
resembles an acyclic hypergraph. 
Hypergraphs of bounded treewidth have also bounded
degree of cyclicity, but not vice-versa. Queries of bounded 
degree of cyclicity can be recognized and processed in polynomial 
time~\cite{gyss-etal-94}.

\item {\em Queries of bounded query-width.} This notion is based on the 
concept of {\em query decomposition}~\cite{chek-raja-97}. 
Any hinge-tree  decomposition of width $k$ is also a query decomposition of 
width $k$, but not vice-versa. Thus query decompositions are the most 
general (i.e., most liberal) decompositions. 
It follows from results in~\cite{chek-raja-97} that queries of
bounded query-width can be answered in polynomial time, once 
a query decomposition is given.
\end{itemize}

Thus, the class of queries of bounded query-width is the widest
of the above mentioned classes of tractable cyclic queries.
We next show that this class is properly included in the class of
queries of bounded hypertree-width. More precisely, we show that every
$k$-width query-decomposition corresponds to an equivalent $k$-width
hypertree-decomposition, but the converse is not true, in general.
Recall that $hw(Q)$ and $qw(Q)$ denote the hypertree width and
the query width of a conjunctive query $Q$.

\begin{theo}\label{theo:HDvsQD}
\ \\ 
\begin{itemize}
\vspace{-0.6cm}
\item[a)] For each conjunctive query $Q$ it holds that $hw(Q)\leq qw(Q)$.
\item[b)] There exist queries $Q$ such that $hw(Q)<qw(Q)$.
\end{itemize}
\end{theo}
\begin{proof} 
({\em Point a.}) Let $Q$ be a conjunctive query and 
$\tuple{T,\lambda}$ a query decomposition of $Q$. W.l.o.g., assume $Q$ is
pure (i.e., labels contain only atoms, see Section~\ref{sec:query-width}).
Then, $(T,\chi,\lambda)$ is a hypertree decomposition of $Q$, 
where, for any vertex $v$ of $T$, $\chi(v)$ consists of the set
of variables $var(\lambda(v))$ occurring in the atoms $\lambda(v)$.
Indeed, because the properties of query decompositions holds for
$\tuple{T,\lambda}$,
$\tuple{T,\chi,\lambda}$ verifies Condition 1 and 2 of Definition
\ref{defn:hyper-decomp}. Condition 3 and 4 follows immediately,
as $\chi(p)=var(\lambda(p))$ by construction.
Therefore, $hw(Q) \leq qw(Q)$.

({\em Point b.}) The query $Q_4$ of Example~\ref{ex:decomposition}
has no query decompositions of width 2, 
and it is easy to see that $qw(Q_4)=3$.
However, $hw(Q_4)=2$, as witnessed by the hypertree decomposition
shown in Figure~\ref{fig:HDvsQD}. 
\end{proof}

\subsection*{Acknowledgment}
We thank Cristinel Mateis for his useful comments on an earlier version
of this paper.


\begin{thebibliography}{99}

\bibitem{abit-etal-95}
S. Abiteboul, R. Hull, V. Vianu.
{\em Foundations of Databases}, Addison-Wesley Publishing Company, 1995.

\bibitem{beer-etal-83}
C. Beeri, R. Fagin, D. Maier, and M. Yannakakis.
On the Desiderability of Acyclic Database Schemes.
{\em Journal of ACM}, 30(3):479--513, 1983.

\bibitem{bern-good-81}
P.A. Bernstein, and N. Goodman.
The power of natural semijoins.
{\em SIAM J. Computing}, 10(4):751--771, 1981.
 
\bibitem{bodl-96}
H. L. Bodlaender. 
A linear-time algorithm for finding tree-decompositions of small treewidth. 
{\em SIAM Journal on Computing}, 25(6):1305-1317, 1996.

\nop{
\bibitem{bodl-97}   
H.L. Bodlaender. 
Treewidth: Algorithmic Techniques and Results. 
In {\em Mathematical Foundations of Computer Science (MFCS'97)},
Bratislava, Slovakia. 
Lecture Notes in Computer Science, Vol. 1295, Springer, pp. 19-36, 1997.  
 
\bibitem{bodl-hage-98} 
H.L. Bodlaender and T. Hagerup.
Parallel Algorithms with Optimal Speedup for Bounded Treewidth.
{\em SIAM J. Computing}, 27(6): 1725-1746, 1998. 
}

\bibitem{chan-etal-81}
A.K. Chandra, D.C. Kozen, and L.J. Stockmeyer.
Alternation.
{\em Journal of ACM}, 26:114--133, 1981.

\bibitem{chan-merl-77}
A.K. Chandra and P.M. Merlin.
Optimal Implementation of Conjunctive Queries in relational Databases.
In In {\em ACM Symp. on Theory of Computing (STOC'77)}, pp.77--90, 1977.


\bibitem{chek-raja-97}
Ch.~Chekuri and A.~Rajaraman.
Conjunctive Query Containment Revisited.
In {\em Proc. International Conference on Database Theory 1997
(ICDT'97), Delphi, Greece, Jan.~1997}, 
Springer LNCS, Vol.~1186, pp.56--70, 1997.


\bibitem{datr-mosc-86}
A. D'Atri, and M. Moscarini.
Recognition algorithms and design methodologies  for acyclic database
schemes.
In {\em Advances in Computing Research}, 3:43--68,
P. Kannellakis, JAI press, London, 1986. 


\bibitem{fagi-83}
R. Fagin.
Degrees of acyclicity for hypergraphs and relational database schemes.
{\em J. ACM}, 30(3):514--550, 1983.

\bibitem{fagi-etal-82}
R. Fagin, A.O. Mendelzon, and J.D. Ullman.
A simplified universal relation assumption and its properties.
{\em ACM Trans. on Database Systems}, 7(3):343--360, 1982.

\bibitem{gare-john-79}
M.R. Garey and D.S. Johnson.
{\em Computers and Intractability. A Guide to the Theory of \NP-completeness}.
Freeman and Comp., NY, USA, 1979.


\bibitem{good-shmu-82}
N. Goodman, and O. Shmueli.
Tree queries: a simple class of relational queries.
{\em ACM Trans. on Database Systems}, 7(4):653--677, 1982.


\bibitem{gott-etal-98}
G.~Gottlob, N.~Leone, and F.~Scarcello.
The Complexity of Acyclic Conjunctive Queries.
Technical Report DBAI-TR-98/17, available on the web as: 
{\tt http://www.dbai.tuwien.ac.at/staff/gottlob/acyclic.ps},
or by email from the authors.
An extended abstract concerning part of this work has been published in
{\em Proc. of the IEEE Symposium on 
Foundations of Computer Science (FOCS'98)}, pp.706--715,
Palo Alto, CA, 1998.

\bibitem{gott-etal-98c}
G.~Gottlob, N.~Leone, and F.~Scarcello.
Computing LOGCFL Certificates.
Technical Report DBAI-TR-98/19, available on the web as:
{\tt http://www.dbai.tuwien.ac.at/staff/gottlob/certificates.ps},
or by email from the authors.

\bibitem{gott-etal-98d}
G.~Gottlob, N.~Leone, and F.~Scarcello.
Hypertree Decompositions and Tractable Queries.
Technical Report DBAI-TR-98/21, available on the web as:
{\tt http://www.dbai.tuwien.ac.at/staff/gottlob/hypertrees.ps},
or by email from the authors.

\bibitem{grei-73}
S.H. Greibach.
The Hardest Context-Free Language.
{\em SIAM Journal on Comput.}, 2(4):304--310, 1973.

\bibitem{gyss-etal-94}
M. Gyssens, P.G. Jeavons, and D.A. Cohen.
Decomposing constraint satisfaction problems using database
techniques.
{\em Artificial Intelligence}, 66:57--89, 1994.

\bibitem{gyss-para-84}
M. Gyssens, and J. Paradaens.
A Decomposition Methodology for Cyclic Databases.
In {\em Advances in Database Theory}, volume 2, pp. 85-122.
Plenum Press New York, NY, 1984.

\bibitem{john-90}
D.~S. Johnson.
\newblock {A Catalog of Complexity Classes}.
\newblock In J.~{van Leeuwen}, editor, {\em {Handbook of Theoretical
Computer
  Science}}, volume~A, chapter~2, pp.67--161. Elsevier Science Publishers B.V.
  (North-Holland), 1990.

\bibitem{kola-vard-98}
Ph.~G.~Kolaitis andf M.~Y.~Vardi.
Conjunctive-Query Containment and Constraint Satisfaction.
To appear in: 
{\em Proc. of Symp. on Principles of Database Systems (PODS'98)},
1998.

\bibitem{maie-86}
D. Maier.
{\em The Theory of Relational Databases}, 
Rochville, Md, Computer Science Press, 1986.

\bibitem{papa-yann-97}
C.H. Papadimitriou and M. Yannakakis.
On the Complexity of Database Queries.
In {\em Proc. of Symp. on Principles of Database Systems (PODS'97)}, 
pp.12--19, Tucson, Arizona, 1997.

\bibitem{qian-96}
X.~Qian.
Query Folding.
In {\em Proc. Internatl. Conf. on Data Engineering,
New Orleans, Louisiana, Feb.26-March  1, 1996 (ICDE'96 }, 
pp.48--55, 1996.

\bibitem{robe-seym-86}
N. Robertson and P.D. Seymour.
Graph Minors II. 
Algorithmic Aspects of Tree-Width. 
{\em J. Algorithms}, 7:309-322, 1986.

\bibitem{ruzz-80}
W.L. Ruzzo.
Tree-Size Bounded Alternation.
{\em Journal of Computer and System Sciences}, 21:218--235, 1980.

\bibitem{skyu-vali-85}
S. Skyum and L.G. Valiant.
A complexity theory based on Boolean algebra.
{\em J. Assoc. Comput. Mach.}, 32:484--502, 1985.

\bibitem{sudb-77}
I.H. Sudborough.
Time and Tape Bounded Auxiliary Pushdown Automata.
In {\em Mathematical Foundations of Computer Science 1977},
LNCS 53, Springer-Verlag, pp.493--503, 1977.

\bibitem{tarj-yann-84}
R.E. Tarjan, and M. Yannakakis.
Simple linear-time algorithms to test chordality of graphs, test
acyclicity
of hypergraphs, and selectively reduce acyclic hypergraphs.
{\em SIAM J. Computing}, 13(3):566-579, 1984.

\bibitem{ullm-89}
J.D. Ullman.
{\em Principles of Database and Knowledge Base Systems, Vol II},
Computer Science Press, Rockville, MD, 1989.

\bibitem{ullm-97}
J.D. Ullman.
Information integration using logical views.
In {\em Database Theory - ICDT'97}, LNCS 1186, pp. 19--40, Springer-Verlag, 
1997.

\bibitem{vard-82}
M. Vardi.
Complexity of Relational Query Languages.
In {\em Proc. of 14th ACM STOC}, pp. 137--146, 1982.

\bibitem{yann-81}
M. Yannakakis.
Algorithms for Acyclic Database Schemes.
In {\em Proc. of Int. Conf. on Very Large Data Bases (VLDB'81)}, pp. 82--94,
C. Zaniolo and C. Delobel Eds., Cannes, France, 1981.


\bibitem{yu-ozso-84}
C.T. Yu, and M.Z. Ozsoyoglu.
On determining tree-query membership of a distributed query.
{\em Infor}, 22(3):261--282, 1984.


\end{thebibliography}
\end{document}